\documentclass[runningheads,orivec]{llncs}
\usepackage{orcidlink}
\usepackage{bbding}
\usepackage[T1]{fontenc}
\usepackage{xcolor}
\usepackage{amssymb}
\usepackage{amsmath}
\usepackage{paralist}
\usepackage{todonotes}
\usepackage{xspace}
\usepackage{pifont}
\usepackage{tikz}
\usetikzlibrary{shapes, shapes.callouts, calc, positioning}
\usetikzlibrary{arrows.meta}
\usepackage{silence} 
\usepackage{thmtools}
\usepackage{thm-restate}

\let\llncssubparagraph\subparagraph
\let\subparagraph\paragraph
\usepackage[compact]{titlesec}
\let\subparagraph\llncssubparagraph
\titlespacing*{\subsubsection} {0pt}{5pt}{2pt} % Acknowledgments, etc.

\usepackage{silence}
\ActivateWarningFilters[captionw]
\usepackage[skip=2pt]{caption}
\DeactivateWarningFilters[captionw]

\usepackage[section]{placeins}
\usepackage{booktabs}
\usepackage{array}

\usepackage{algorithm}
\usepackage[noend]{algpseudocode}
\usepackage{algorithmicx}
\usepackage[shortlabels,inline]{enumitem}
\usepackage[capitalise]{cleveref}
\crefname{algorithm}{alg.}{algs.}

\newif\ifextended
\extendedtrue % Change this to build the extended version
\newcommand{\extended}[1]{\ifextended #1 \fi}
\newcommand{\reduced}[1]{\ifextended\else #1 \fi}

\newcommand{\algname}[0]{Data-Aware Declare Aligner}
\newcommand{\toolname}[0]{DADA}
\newcommand{\declare}[0]{Declare\xspace}
\newcommand{\smt}[0]{SMT}
\newcommand{\smtlib}[0]{SMT-LIB2}

\newcommand{\m}[1]{\mathsf{#1}}
\newcommand{\Dom}{\mathcal D} % domain of sorts
\newcommand{\bool}{\mathtt{bool}} % sorts
\newcommand{\integer}{\mathtt{int}}
\newcommand{\str}{\mathtt{string}}
\newcommand{\rational}{\mathtt{rat}}
\newcommand{\seq}[2][n]{{#2_1},\dots,{#2_{#1}}} % sequence
\newcommand{\tup}[1]{\langle#1\rangle}
\newcommand{\dom}{\mathit{dom}} % domain of partial function
\newcommand{\varinst}[2]{[#1](#2)} % instantiation of constraint to event variables
\newcommand{\eid}[1]{\texttt{\#}_{#1}} % event id

\newcommand{\CC}{\mathcal C} % set of all constraints
\newcommand{\Act}{\mathcal A} % set of activities
\newcommand{\Id}{\mathit{Id}} % set of event ids
\newcommand{\trace}{\mathbf e} % log trace
\newcommand{\E}{E} % some set of events
\newcommand{\run}{\mathbf f} % model run (or candidate thereof)
\newcommand{\events}{E} % set of all events
\newcommand{\SKIP}{{\gg}}% skip move in alignment
\newcommand{\eventsplus}{\events^\SKIP} % set of events incl SKIP
\newcommand{\moves}{\mathit{Moves}} % set of moves
\newcommand{\restr}[2]{\left.#1\right|_{#2}} % restriction
\newcommand{\runsof}[1]{\mathit{runs}(#1)} % runs satisfying DECLARE model
\newcommand{\alignments}[1]{\mathit{Align}(#1)} % alignments for trace
\newcommand{\MM}{\mathcal M}
\newcommand{\cost}{\mathit{cost}} % cost of A* state
\newcommand{\Ord}{\mathit{Ord}} % order constraints of a declare constraint
\newcommand{\dirbefore}{\mathrel{\ll}} % directly before
\newcommand{\altbefore}[2]{<^{#1}_{[#2]}} % alternate/conditionally before
\newcommand{\tgt}{\mathit{tgt}}
\newcommand{\act}{\mathit{act}}
\newcommand{\cor}{\mathit{cor}}
\newcommand{\cA}{c_\act}
\newcommand{\cT}{c_\tgt}
\newcommand{\cR}{c_\cor}

\newcommand{\DC}[1]{\mathrm{#1}} % formatting of declare constraint

\newcommand{\syncmove}[4]{\tikz[baseline=-0.5ex]{\node[scale=.8, inner sep=0pt]{$%
\begin{array}{|@{\,}l@{\quad}l@{\,}|}%
\hline{#1}&{#2}\\\hline{#3}&{#4}\\\hline\end{array}$}}}
\newcommand{\logmove}[2]{\tikz[baseline=-0.5ex]{\node[scale=.8, inner sep=0pt]{$%
\begin{array}{|@{\,}l@{\quad}l@{\,}|}%
\hline{#1}&{#2}\\\hline\multicolumn{2}{|c|}{\SKIP}\\\hline\end{array}$}}}
\newcommand{\modelmove}[2]{\tikz[baseline=-0.5ex]{\node[scale=.8, inner sep=0pt]{$%
\begin{array}{|@{\,}l@{\quad}l@{\,}|}%
\hline\multicolumn{2}{|c|}{\SKIP}\\\hline{#1}&{#2}\\\hline\end{array}$}}}

\tikzstyle{follows}=[->, thick ,>={Stealth}]
\tikzstyle{dirfollows}=[->, thick,double, >={Stealth}]

\newcommand{\tmp}{\tau} % timestamp variable

\reduced{\renewcommand{\todo}[2][]{}} % Include comments only in extended for now...

\begin{document}

\title{Efficient Conformance Checking of Rich Data-Aware Declare Specifications\extended{ (Extended)}
% OLD VERSION \algname{}: Efficient SMT-based Multi-Perspective Declarative Conformance Checking
}
\titlerunning{Efficient Conformance Checking of Rich Data-Aware Declare Specifications}
\author{Jacobo Casas-Ramos\inst{1}\Envelope\orcidlink{0000-0003-2521-1464}
\and Sarah Winkler\inst{2}\orcidlink{0000-0001-8114-3107}
\and Alessandro Gianola\inst{3}\orcidlink{0000-0003-4216-5199}
\and Marco Montali\inst{2}\orcidlink{0000-0002-8021-3430}
\and Manuel Mucientes\inst{1}\orcidlink{0000-0003-1735-3585}
\and Manuel Lama\inst{1}\orcidlink{0000-0001-7195-6155}
}
\authorrunning{Casas-Ramos, J. et al.}
\institute{Universidade de Santiago de Compostela, Santiago de Compostela, Spain \email{\{jacobocasas.ramos,manuel.lama,manuel.mucientes\}@usc.es} \and
Free University of Bozen-Bolzano, Bolzano, Italy
\email{\{montali,winkler\}@inf.unibz.it} \and
INESC-ID/Instituto Superior Técnico, Universidade de Lisboa, Lisbon, Portugal
\email{alessandro.gianola@tecnico.ulisboa.pt}}

\maketitle

\begin{abstract}
Despite growing interest in process analysis and mining for data-aware specifications, alignment-based conformance checking for declarative process models has focused on pure control-flow specifications, or mild data-aware extensions limited to numerical data and variable-\-to-constant comparisons. This is not surprising: finding alignments is computationally hard, even more so in the presence of data dependencies. In this paper, we challenge this problem in the case where the reference model is captured using data-aware Declare with general data types and data conditions. We show that, unexpectedly, it is possible to compute data-aware optimal alignments in this rich setting, enjoying at once efficiency and expressiveness. This is achieved by carefully combining the two best-known approaches to deal with control flow and data dependencies when computing alignments, namely A* search and SMT solving. Specifically, we introduce a novel algorithmic technique that efficiently explores the search space, generating descendant states through the application of repair actions aiming at incrementally resolving constraint violations. We prove the correctness of our algorithm and experimentally show its efficiency. The evaluation witnesses that our approach matches or surpasses the performance of the state of the art while also supporting significantly more expressive data dependencies, showcasing its potential to support real-world applications.
\keywords{Multi-perspective conformance checking \and Efficient optimal alignments \and Data-aware \declare{} \and Satisfiability modulo theories (SMT).}
\end{abstract}

\section{Introduction}

% In the realm of process mining, conformance checking has gained traction \cite{PMH2022} as one of its main analytic tasks. It consists of detecting deviations and discrepancies between the observed behavior of a process, given by a sequence of distinct events recorded during a process execution, and its expected behavior defined by a reference process model.

% In conformance checking, process models can be represented using either procedural or declarative approaches. Procedural models explicitly define all admitted sequences of activities, providing a direct representation of the expected behaviors of the process. In contrast, declarative models offer a set of constraints and rules that regulate the acceptable process behaviors, without explicitly enumerating specific execution sequences. The most famous declarative modeling formalism is given by \declare{} \cite{CiccioM22}, whose formal semantics is defined by leveraging Linear Time Logic over finite traces (\textsc{ltl}$_f$) \cite{DGV13}.

% 

Conformance checking~\cite{CarmonaDSW18} is a cornerstone task in process mining. % The conformance checking task 
It relates the observed behaviour contained in an event log to the expected behaviour described by a reference process model, with the goal of identifying and reporting deviations. A widely adopted approach substantiates conformance checking in the computation of so-called optimal \emph{alignments}, where each non-conforming log trace is compared against the closest model trace(s), indicating where discrepancies are located and calculating a corresponding cost %, typically using variants of string edit distances} 
\cite{BosA12}. 

Lifting the computation of alignments to process models integrating multiple perspectives (most prominently data and control-flow) has been tackled with increasing interest \cite{MannhardtLRA16,Bergami2021,FelliGMRW21,Felli2023}, and so has been dealing with other forms of data-aware conformance checking. On the one hand, this reflects a growing prominence of multi-perspective (in particular, data-aware) process models in the foundations of process science. On the other hand, computing multi-perspective alignments provides more informative insights than pure control-flow alignments \cite{MannhardtLRA16}. 

Despite this growing interest, alignment-based conformance checking for declarative data-aware process specifications is still an open problem. Existing work mainly focuses only on control-flow, concretely expressed using DCR graphs \cite{ChrS23}, \declare{} \cite{DBLP:journals/is/LeoniMA15}, or Linear Temporal Logic on finite traces (LTLf) \cite{DeGiacomoMMP17}.
To the best of our knowledge, the only attempt of lifting alignment computation to a data-aware setting is \cite{Bergami2021}, which considers however a very limited data-aware extension of \declare{} where data are numerical and data conditions are restricted to variable-to-constant comparisons, such as $x > 5$,
which also excludes comparisons between variables in different events. 
% The need for considering richer multi-perspective dependencies in constraints,\footnote{As it will become clear in the paper, we treat timestamps as a specific datatype, hence data constraints also cover quantitative time constraints, such as deadlines.} is well-known (see, e.g., \cite{BurattinMS16,MannhardtLRA16,FelliGMRW21}), and exemplified next.

\begin{example}
To highlight the 
%flexibility of \smt{}-based 
expressivity of complex
data conditions, consider a model for a shipping company that has a \declare{} \emph{response} constraint between two events: ``Package Shipment'' (A) and ``Delivery Confirmation'' (B), such that after a package is shipped, delivery confirmation must be received. The constraint is equipped with a data condition that specifies that the delivery confirmation must be received within 3 days of shipping if the package weighs less than 10 kg and the delivery address is within a specific geographic region. In any other case, it must be received within 10 days of shipping. Our approach can handle this condition, expressed in \smtlib{}~\cite{Barrett2025} syntax as \texttt{(A.weight < 10.0 and B.region == "Europe") ? (B.time - A.time <= 3d) : (B.time - A.time <= 10d)}. %using the concrete syntax of \smtlib{}.
\end{example}

The fact that previous work did not consider complex constraints like the one shown in the example is not surprising: finding alignments is computationally hard, even more so in the presence of data dependencies. Also, this more expressive setting cannot be addressed relying on previous methods \cite{DeGiacomoMMP17,Bergami2021}, based on the construction of an automaton capturing all and only the traces accepted by the reference \declare{} specification. In fact, this is not possible even for numerical datatypes going beyond mere comparison predicates, due to undecidability \cite{FMPW23}.

% In this paper, we tackle this challenging problem by introducing a foundational algorithmic approach, paired with an effective implementation, to compute alignments for rich data-aware \declare{} specifications. Constraints are extended with expressive data conditions (both local to time instants and correlating time instances) over various datatypes, ranging from primitive datatypes such as strings and numbers to full-fledged data structures \cite{Gianola2023}---essentially, all types %and conditions 
% supported by state-of-the-art satisfiability modulo theory (\smt{}) solvers \cite{Moura2008,Dutertre2014}.

In this paper, we tackle this challenge by casting the alignment problem as a search problem that is solved by repeatedly identifying and repairing constraint violations.
%, thereby intervening on model, trace, edit, and synchronous moves. 
% This requires to reason on time and data conditions at once to explore the space of possible repairs efficiently.
To this end we strategically integrate the two most effective methods for  handling control flow and data dependencies in alignment computation, respectively: A$^*$ search, in the variant adopted by one of the most recent methods for \declare{} \cite{CasasRamos2025}, and \smt{} solving \cite{Barrett2025}, so far employed for aligning data-aware procedural process models~\cite{FelliGMRW21}. %,Felli2023}. 
Our technique defines a novel search space that is explored using the A$^*$ search algorithm to find an optimal alignment. The initial state of the search space represents the original trace as an \smt{} formula. When a state is explored, an \smt{} solver is used to identify constraint violations, which in turn triggers the generation of child states by repairing the parent state. This process continues until a goal state is found that has minimal cost and no violation to repair, from which the optimal alignment can be reconstructed.

We establish correctness of our algorithm through rigorous proofs and provide an extensive experimental evaluation, showing its ability to operate efficiently even when complex data conditions are employed. Our method matches or surpasses the performance of the state of the art while providing for the first time concrete support for rich datatypes and data conditions.

\smallskip
The remainder of this paper is structured as follows:
Sec.~\ref{sec:related} we summarize related work.
In Sec.~\ref{sec:preliminaries} we recall the necessary preliminaries about data-aware \declare, and alignments. 
Sec.~\ref{sec:dada} is dedicated to our approach to conformance checking for data-aware \declare specifications.
We describe its implementation in the tool {\toolname} in Sec.~\ref{sec:implementation}.
In Sec.~\ref{sec:evaluation}, we provide a detailed evaluation and comparison with the state-of-the-art.
In Sec.~\ref{sec:conclusions}, we conclude and give some directions for future work.
Additional material is available online\footnote{\url{https://apps.citius.gal/dada} and \url{https://doi.org/10.5281/zenodo.15470077}}\reduced{ and in an extended version of this paper~\cite{extended}\ignorespaces}.

\section{Related Work}
\label{sec:related}

Although significant research has focused on computing optimal alignments for imperative process models that incorporate the data perspective~\cite{MannhardtLRA16,FelliGMRW21,nagy2022alignment,de2013aligning}, there is a notable lack of work on data-aware alignment techniques for declarative models.
Existing conformance checking methods for declarative models primarily focus on control-flow \cite{LeoniMA12,DeGiacomoMMP17,GiacomoFMMP23,CasasRamos2025,ChrS23,DBLP:journals/is/LeoniMA15}. Notably, they lack the capability to handle data conditions, a critical component of real-world processes. %Our approach addresses this gap by integrating SMT solving for comprehensive handling of both temporal and data conditions, guided by A* search to find an optimal alignment.
Though some approaches have been proposed for conformance checking of data-aware \declare{} models, they have severe limitations. 
SQL queries have been employed to filter traces from a database that match the given specification, but this approach only considers exactly matching traces ~\cite{Schonig2016,RivaBMMM23}. 
Similarly, \cite{Borrego014} uses constraint programming for trace analysis, but only supports global data. 
More comprehensive results are provided by the analysis framework
\cite{BurattinMS16}, which reports activations, fulfillments, and violations of constraints. 
Moreover, importantly, none of the approaches~\cite{Schonig2016,RivaBMMM23,Borrego014,BurattinMS16} provides alignments, thus failing to offer detailed insights into the nature and extent of deviations between observed and expected behavior:
% instead relying on simplistic metrics such as the number of constraint violations. 
This lack of nuanced analysis hinders the ability to identify root causes of non-conformance and implement targeted improvements, ultimately undermining the effectiveness of conformance checking efforts.

Closest to our work is the planning-based conformance checking approach \cite{Bergami2021}, which does compute optimal alignments of data-aware \declare{} models. However, their treatment of the data perspective has severe limitations: data conditions can only refer to activation or target events, excluding correlation conditions that link the two. Moreover, data conditions are restricted to simple variable-to-constant comparisons, whereas our approach supports much more expressive data conditions over
%thanks to the use of SMT solvers as backends. These in 
a wide range of data types including integers, bit-vectors, infinite-precision reals, and arrays. Specifically, we support the complete language of conditions specified in the \smtlib{} standard~\cite{Barrett2025}, requiring only that the underlying {\smt} theory is decidable.
Overall, our approach offers a significant improvement over existing methods in terms of expressiveness and efficiency. % repeated: making it a valuable tool for aligning complex processes while maintaining computational efficiency and expressive richness.
% \sarahtodo{reformulated a bit}

\section{Preliminaries}
\label{sec:preliminaries}

In this section we introduce the required background about event logs, \declare with data, and alignments. We start with the data condition language and events.

\paragraph{Data conditions.}
% \jacobotodo{Moved data conditions before event to define $V$. Sarah: good}
We consider \emph{sorts} $\Sigma = \{\texttt{bool}, \texttt{int}, \texttt{rat},\texttt{string}\}$ for data payloads, with associated domains 
$\Dom(\bool) = \mathbb B$, the booleans;
$\Dom(\integer) = \mathbb Z$, the integers;
$\Dom(\rational) = \mathbb Q$, the rational numbers, and 
$\Dom(\str) = \mathbb S$, finite strings.
For a set of variables $V$ and a sort $\sigma\in \Sigma$, 
$V_{\sigma}$ denotes the subset of $V$ of sort $\sigma$. 

\begin{definition}
\label{def:data:conditions}
 A \emph{data condition} over a set of variables $V$ is an expression $c$ according to the following grammar:\\
%
%\begin{align*}
% $
% \begin{array}{@{}rl@{\quad}rl@{}}
%  c &= V_{\bool} \mid \mathbb B \mid n \geq n \mid r \geq r \mid r > r \mid s = s \mid b \wedge b \mid \neg b &
%  s &= V_{\str} \mid \mathbb S \\
%  n &= V_{\integer} \mid \mathbb Z \mid n + n \mid - n &
%  r &= V_{\rational} \mid \mathbb Q \mid r + r \mid - r
% \end{array}
% $\\
$
\begin{array}{@{}rl@{\quad}rl@{}}
 c &:= v_{\bool} \mid b \mid n \geq n \mid r \geq r \mid r > r \mid s = s \mid c \wedge c \mid \neg c &
 s &:= v_{\str} \mid t \\
 n &:= v_{\integer} \mid z \mid n + n \mid - n &
 r &:= v_{\rational} \mid q \mid r + r \mid - r
\end{array}
$\\
%\end{align*}
%
where $v_\sigma \in V_\sigma$ for $\sigma \in \Sigma$,
$b \in \mathbb B$,
$t \in \mathbb S$,
$z \in \mathbb Z$, and
$q \in \mathbb Q$.
% \sarahtodo{changed to avoid sets on rhs, which is unusual}
The set of data conditions over a set of variables $V$ is denoted by $\CC(V)$.
\end{definition}

We use data conditions as in Def.~\ref{def:data:conditions} in this paper to have a concrete language to refer to, but our implementation actually allows for 
%the full 
arbitrary conditions in the
\smtlib{} language~\cite{Barrett2025} that is supported by the \smt{} solver of choice.
% \jacobotodo{Added decidability comment. Sarah: proposal for reform using common terminology of 'SMT theory being decidable'}
In the sequel, we assume that the underlying SMT theory is decidable, though; this restriction is required to provide correctness guarantees. 
%Note that our theoretical results rely on a decidable subset of \smtlib{}, which provides a foundation for our proofs and analysis.
%One can thus also e.g. use data conditions with uninterpreted functions and predicates; or bitvector expressions. % (Examples of complex conditions are already given in the evaluation)

\paragraph{Event logs.}
Below, we assume an arbitrary infinite set $\Id$ of event identifiers; and a set $\Act$ of activities, where elements $a \in \Act$ are denoted by lower-case letters.
% \sarahtodo{moved first mention of $\Act$ here}
We consider the following notions of data-aware events, traces, and event logs:

\begin{definition}
An \emph{event} $e$ is a triple $e=(\iota, a,\alpha)$ such that $\iota\in \Id$, $a\in \Act$ is an activity,
and $\alpha$ is a partial assignment that maps variables in $V$ to elements of their domain.
Given a set of events $E$, a \emph{trace} $\trace$ is a finite sequence of events in $E$, that is, $\trace \in E^*$; and
an \emph{event log} is a multiset of traces.
The domain of an assignment $\alpha$ is denoted $\dom(\alpha)$.
% \sarahtodo{added as it is used below and was never introduced. Jacobo: merged with definition to save space.}
\end{definition}

\paragraph{\declare.}
% In the sequel, we assume that $\Act$ is a fixed set of activities, denoted by lower-case letters. 
Tab.~\ref{tab:declare} lists the \declare templates used in this paper. %\footnote{Our implementation supports in fact the full list in~\cite[Tab.~2]{CasasRamos2025}.}
We call a \emph{\declare constraint} an expression that is obtained from a \declare template by substituting the upper-case template variables by activities in $\Act$. 
Constraints based on templates (6)--(9) and (10)--(12) are called \emph{response} and \emph{precedence constraints}, respectively, while constraints using (13)--(15) are \emph{negation constraints}.

\begin{table}[t]
\centering
\begin{scriptsize}\setlength\extrarowheight{-1pt}
\begin{tabular}{@{}l@{\,}l@{\,}l@{}}
\bfseries (1) & $\DC{Existence}(n, A)$:& $A$ occurs at least $n$ times. 	\\
%Q activation \top or A? Rather \top, so that missing A is missing target?
%Participation(A): A occurs at least once.
\bfseries (2) & $\DC{Absence}(n, A)$: & $A$ occurs at most $n-1$ times. \\		
%Q violation counted as?
%AtMostOne(A): A occurs at most once.
%Exactly(n, A): A occurs exactly n times. 
\bfseries (3) & $\DC{Init}(A)$: & $A$ is the first activity.\\
\bfseries (4) & $\DC{End}(A)$: & $A$ is the last activity.  \\
\bfseries (5) & $\DC{Choice}(A, B)$: &Either $A$ or $B$, or both, occur. \\
% (6) & $\DC{ExclusiveChoice}(A, B)$: & Either $A$ or $B$, but not both, occur. \\
\bfseries (6) & $\DC{RespondedExistence}(A, B)$: & If $A$ occurs, $B$ also occurs. \\
\bfseries (7) & $\DC{Response}(A, B)$: & If $A$ occurs, $B$ follows. \\
\bfseries (8) & $\DC{AlternateResponse}(A, B)$: & If $A$ occurs, $B$ follows without an $A$ in between. \\
\bfseries (9) & $\DC{ChainResponse}(A, B)$: & If $A$ occurs, $B$ is the next activity. \\
\bfseries (10) & $\DC{Precedence}(A, B)$: & If $B$ occurs, $A$ precedes it. \\
\bfseries (11) & $\DC{AlternatePrecedence}(A, B)$: & If $B$ occurs, $A$ precedes it without a $B$ in between. \\
\bfseries (12) & $\DC{ChainPrecedence}(A, B)$: & If $B$ occurs, $A$ is the previous activity. \\
%CoExistence(A, B): If A occurs, B also occurs, and vice versa.
%Succession(A, B): Combines Response and Precedence.
%AlternateSuccession(A, B): Combines AlternateResponse and AlternatePrecedence.
%ChainSuccession(A, B): Combines ChainResponse and ChainPrecedence.
%NotCoExistence(A, B): If A occurs, B does not, and vice versa.
%(16) & $\DC{NotPrecedence}(A, B)$: If $B$ occurs, $A$ does not precede it. \\
% (13) & $\DC{NotSuccession}(A, B)$: & If $A$ occurs, $B$ does not follow, \\
% &&and if $B$ occurs, $A$ does not precede it. \\
%(17) & $\DC{NotChainPrecedence}(A, B)$: $B$ is not immediately preceded by $A$. %\\
% (14) & $\DC{NotChainSuccession}(A, B)$: & $A$ is not immediately followed by $B$ \\
% &&and $B$ is not immediately preceded by $A$.
\bfseries (13) & $\DC{NotResponse}(A, B)$: &If $A$ occurs, $B$ does follow. \\
\bfseries (14) & $\DC{NotRespondedExistence}(A, B)$: &If $A$ occurs, $B$ does not. \\
\bfseries (15) & $\DC{NotChainResponse}(A, B)$: &If $A$ occurs, $B$ is not the next activity.
\end{tabular}
\end{scriptsize}
\caption{Supported \declare templates.\label{tab:declare}}
\end{table}

\declare templates, as well as the derived constraints, have \emph{activations} and \emph{targets}.
Intuitively, an activation is an event whose occurrence imposes the (non) occurrence of other events. These other events are called targets.
For the templates in Tab.~\ref{tab:declare}, in all response and negation templates, variable $A$ is the activation and $B$ the target; while in all precedence templates, $B$ is the activation and $A$ the target. 
In the remaining patterns, both $A$ and $B$ are targets.
Given a \declare constraint $\varphi$, an activity is an activation (resp. target) activity in $\varphi$ if it is substituted for an activation (resp. target) variable in the underlying template.

We consider \emph{multi-perspective} \declare constraints that include data conditions.
To that end, for the remainder of the paper we fix a set of sorted \emph{process variables} $V$.
Intuitively, these variables are considered the payload of activities;
they are maintained along the entire trace, but may change their values.
For a set $Set$, let $V^{Set} = \{v_s \mid v\in V\text{ and }s\in Set\}$ be a set of labelled variables that contains a copy of each variable in $V$ for each element in $Set$.
In particular,  
%for $a\in \Act$, the idea is that $v_a$ 
$v_a \in V^\Act$ will be used to represent the value of $v$ while observing activity $a$.

\begin{definition}
\label{def:declare:constraint}
A \emph{\declare constraint with data} is a quadruple $\tup{\varphi,\cA, \cT, \cR}$ consisting of a \declare constraint $\varphi$ and data conditions $\cA, \cT$ and $\cR$.
Precisely, for $a \in \Act$ the activation and 
%$t \in \Act$ the target activity in 
$T \subseteq \Act$ the target activities of
$\varphi$:
\begin{inparaenum}[\itshape (i)]
\item
$\cA \in \CC(V^{\{a\}})$ is called the \emph{activation condition},
\item
%$\cT \in \CC(V^{\{t\}})$ 
$\cT \in \CC(V^{T})$
is called the \emph{target condition}, and 
\item
%$\cR \in \CC(V^{\{a,t\}})$ 
$\cR \in \CC(V^{\{a\} \cup T})$
is the \emph{correlation condition}.
\end{inparaenum}
%\sarahtodo{defined macros \texttt{\textbackslash cA}, \texttt{\textbackslash cT}, \texttt{\textbackslash cR}, which are currently $\cA$, $\cT$, $\cR$ to avoid the confusion that the reviewer complained about, with both $a$ and $A$ being used as a subscript. but feel free to change}
% \sarahtodo{changed as there can be multiple target activities, e.g. for Choice constraints - right? Yes, this is true}
\end{definition}

Intuitively, $\cA$ constrains the data variables while the activation activity is observed, $\cT$ the data variables while the target activity is observed, and $\cR$ expresses relationships between the data variables of both activities. %
For \declare constraints $\varphi$ without activation, we assume that all but $\cT$ are $\top$. 
For simplicity of presentation, we assume that the activation and target activity are different, (though in our implementation this is not required).
A \emph{\declare specification} $\MM$ is a set of \declare constraints with data. In the sequel, if no confusion can arise, we refer to \declare constraints with data simply by \emph{constraints}.

% In examples, we will denote a constraint $(\varphi, \cA, \cT, \cR)$ as $\varphi | \cA | \cT | \cR$, and omit components where the data condition is $\top$.

\begin{example}
\label{exa:running:spec}
As running example, we use the set of variables $V = \{x\}$, activities $\Act = \{\m a, \m b, \m c\}$ and the specification $\MM$ that consists of the following two constraints $\psi_1$ and $\psi_2$, where for readability we write $a.v$  instead of $v_a$, for $a\in \Act$:
\begin{compactitem}
    \item $\psi_1=\tup{\DC{ChainResponse}(\m a, \m c), \top, \top,\m c.x > \m a.x}$: This
    specifies that each occurrence of $\m a$  must be directly followed by an event with $\m c$ such that the value of $x$ associated with activity $\m c$ is greater than the value of $x$ seen with $\m a$.
    \item $\psi_2=\tup{\DC{AlternatePrecedence}(\m c, \m b),\m b.x \ge 0,\m c.x \neq 0,\m c.x < \m b.x}$:
    This states an alternate precedence relationship between the activation $\m b$ and the target $\m c$, demanding that if the value of $x$ seen with $\m b$ is non-negative, an activity $\m c$ must occur before activity $\m b$, without any other $\m b$ activities with $x \geq 0$ in between. Furthermore, the $x$-value of $\m c$ must be lower than the $x$-value of $\m b$.
\end{compactitem}
\end{example}

The semantics of \declare constraints with data is the same as in~\cite{Bergami2021},
we recall it in~\reduced{\cite[Def. 11]{extended}\ignorespaces}\extended{Sec. \ref{def:declare:constraints}\ignorespaces}.
The set of all traces that satisfy all constraints in $\MM$ is denoted by $\mathit{runs}(\MM)$.
We assume for our approach that $\mathit{runs}(\MM) \neq \emptyset$.

\paragraph{Alignments.}
We aim to design a conformance-checking procedure that, given a trace and a \declare specification $\MM$, finds an optimal alignment of $\trace$ and a run of $\MM$. 
Typically, when constructing alignments, not all events in the trace can be put in correspondence with an event in a run, and vice versa. 
Hence we use a  ``skip'' symbol $\SKIP$ and consider the extended set of events $\eventsplus = \events \cup \{\SKIP\}$.

For a set $\events$ of events, a pair $(e,f)\in {\eventsplus}^2 \setminus \{(\SKIP,\SKIP)\}$ is called \emph{move} iff 
%it is one of:
one of the following cases applies: it is a
\begin{inparaenum}[(i)]
\item \emph{log move} if $e\,{\in}\, \events$ and $f\,{=}\,\SKIP$; 
\item \emph{model move} if $e\,{=}\,\SKIP$ and $f\,{\in}\,\events$; 
\item \emph{edit move} if $(e,f) \in \events^2$, $(e,f)=((\iota,a,\alpha), (\iota,a,\alpha'))$, $\dom(\alpha) = \dom(\alpha')$ and $\exists v \in \dom(\alpha)$ such that $\alpha(v) \neq \alpha'(v)$; 
\item \emph{synchronous move} if $(e,f) \in \events^2$ and $e = f$.
\end{inparaenum}
We denote by $\moves$ the set of all moves.

For a sequence of moves 
$\gamma = \langle(e_1,f_1), \dots, (e_n,f_n)\rangle$,
the \emph{log projection} $\restr{\gamma}{L}$ of $\gamma$ is the maximal subsequence $\seq[i]{e'}$ of $\seq[n]e$ such that $\seq[i]{e'}\in \events^*$, that is, it contains no $\SKIP$ symbols.
Similarly, the \emph{model projection} $\restr{\gamma}{M}$ of $\gamma$ is the maximal subsequence $\seq[j]{f'}$ of $\seq[n]f$ such that $\seq[j]{f'}\in \events^*$.

\begin{definition}[Alignment]
Given a \declare model $\MM$, a sequence of moves $\gamma$
is an \emph{alignment} of a trace $\trace$ against $\MM$ if $\restr{\gamma}{L} = \trace$, and $\restr{\gamma}{M}\in\runsof{\MM}$.
The set of alignments for a trace $\trace$ wrt. $\MM$ is denoted by $\alignments{\MM, \trace}$.
\end{definition}

\begin{example}
\label{exa:running:alignments}
Consider the trace
$\trace=\langle(\eid{1}, \m a, \{x=0\}), (\eid{2}, \m b, \{x=2\})\rangle$.
The following are two possible alignments for $\trace$ against the model from Ex.~\ref{exa:running:spec}:
\[
\gamma_1=
\syncmove{\m a}{\{x=0\}}{\m a}{\{x=0\}}
\modelmove{\m c}{\{x=1\}}
\syncmove{\m b}{\{x=2\}}{\m b}{\{x=2\}}
\qquad
\gamma_2=
\syncmove{\m a}{\{x=0\}}{\m a}{\{x=0\}}
\modelmove{\m c}{\{x=3\}}
\logmove{\m b}{\{x=2\}}
\]
Each move $(e, f)$ is shown in a column, including $e$ in the first row and $f$ in the second row. Since event identifiers are irrelevant in alignments, we omit them.
\end{example}

A \emph{cost function} is a mapping $\kappa\colon \moves \to \mathbb R^+$ that assigns a cost to every move. It is naturally extended to alignments as follows. 

\begin{definition}[Alignment cost]
\label{def:cost}
Given %$\MM$, $\trace$ and 
$\gamma \in \alignments{\MM, \trace}$ as before, the \emph{cost} of $\gamma$ is 
defined as the sum of the costs of its moves, that is,
$\kappa(\gamma) = \sum_{i=1}^n \kappa(e_i,f_i)$.
Moreover, $\gamma$ is
\emph{optimal for $\trace$ and $\MM$} if $\kappa(\gamma)$ is minimal among all alignments for $\trace$ and $\MM$, namely there is no $\gamma'\in \alignments{\MM, \trace}$ with $\kappa(\gamma')<\kappa(\gamma)$.
\end{definition}

\noindent
% We denote the cost of an optimal alignment for $\trace$ with respect to 
% $\MM$ by $\kappa_\MM^{opt}(\trace)$. 
%Given $\MM$, the set of optimal alignments for $\trace$ is denoted by $\OPTalignment{\logtrace}$.

In this paper, we will use the standard cost function $\kappa$ that assigns $\kappa(e,f)=1$ if $(e,f)$ is a log or model move, 
$\kappa(e,f)=0$ if $(e,f)$ is a synchronous move, and for an edit move $(e,f)=((\iota,a,\alpha), (\iota,a,\alpha'))$, $\kappa(e,f)=|\{\alpha(v) \neq \alpha'(v)\mid v\in V\}|$.
% However, our approach supports other cost functions as well, as long as each log and model move and each edit in the payload have a fixed cost. %Implementation detail?

\section{\algname{}}
\label{sec:dada}

In this section, we outline the conceptual approach of the \algname.
Given a \declare specification $\MM$ and a trace $\trace$, the aim is to find an optimal alignment of $\trace$ wrt. $\MM$.
To that end, the basic idea is to start with the event sequence in $\trace$, and subsequently \emph{repair} it until an event sequence is obtained that satisfies all constraints in $\MM$.
To navigate through a large search space of possible repairs and respective alignments while ensuring an optimal solution,
our approach leverages the A$^*$ algorithm.
% As is common in A$^*$, the search is performed on \emph{states} that come with a \emph{cost}.

We use the term \emph{state} to refer to a representation of a candidate alignment.
The formal definition of a state is given below;
intuitively,
%without yet formalizing its structure. This informal introduction facilitates an overview of the algorithm.
% \jacobotodo{Added this explanation to the first mention of state: S: reformulated a bit} 
%Each a \emph{state} 
each state consists of a partially ordered set of events together with data conditions, effectively acting as a candidate alignment which need not yet satisfy all constraints. Moreover, each state has a \emph{cost}, reflecting the cost of alignments extracted from it.
% The cost of each visited state is an underestimate of the cost of the resulting alignment.

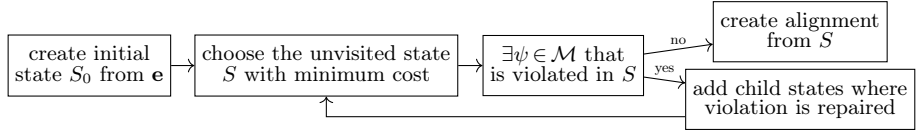
\begin{figure}[t]
% \resizebox{\textwidth}{!}{ % Avoids Underfull \vbox, but wastes more space before caption
\begin{tikzpicture}[node distance=30mm]
\tikzstyle{algstep}=[rectangle,draw,scale=.9, inner sep=3pt]
\tikzstyle{arr}=[->]
\node[algstep] (init) {\begin{tabular}{@{}c@{}}create initial \\[-.6ex] state $S_0$ from $\trace$\end{tabular}};
\node[algstep,right of=init,xshift=5mm] (choose) {\begin{tabular}{@{}c@{}}choose the unvisited state\\[-.6ex]$S$ with minimum cost\end{tabular}};
\node[algstep,right of=choose,xshift=5mm] (viol) {\begin{tabular}{@{}c@{}}$\exists\psi\,{\in}\,\MM$ that\\[-.7ex]is violated in $S$\end{tabular}};
\node[algstep,right of=viol,yshift=-5mm,xshift=5mm] (repair) {\begin{tabular}{@{}c@{}}add child states where\\[-.7ex]violation is repaired \end{tabular}};
\node[algstep,right of=viol,yshift=5mm,xshift=5mm] (align) {\begin{tabular}{@{}c@{}}create alignment\\[-.7ex] from $S$\end{tabular}};
\draw[arr] (init) -- (choose);
\draw[arr] (viol) -- node[above,scale=.6] {no} (align);
\draw[arr] (viol) -- node[above,scale=.6] {yes}  (repair);
\draw[arr] (choose) -- (viol);
\draw[arr] (repair.south west)+(0,5pt) -| (choose);
\end{tikzpicture}
% }
\caption{Overview of the approach.\label{fig:overview}}
\end{figure}

An overview of the approach is sketched in Fig.~\ref{fig:overview}:
the initial state $S_0$ represents the set of events in the input trace, ordered as in $\trace$, and with data conditions that reflect the variable assignments.
The procedure then selects the previously unvisited state $S$ of minimal cost.
It is checked whether there are remaining constraint violations in $S$.
In this case, all possible \emph{repairs} are applied to $S$ creating a new child state from each repair, and another search iteration is performed.
Otherwise, an optimal alignment for $\trace$ is reconstructed from $S$.

\subsection{State definition}
% We start by explaining in detail our notion of a \emph{state}.
As mentioned above, a state contains a partially ordered set of events, and data conditions on their payloads. In order to express conditions that involve variables in all events, we need, as a technicality, labelled variables:
For an event $e=(\iota,a,\alpha)$, let $V^e = \{v_\iota \mid v\in V\}$ be a copy of the set $V$ where each variable is labelled by the id of $e$.
For a set of events $\E$, let $V^\E =\bigcup_{e \in \E} V^e$ be the set of variables for all events in $\E$. A state with set of events $\E$ can then use data conditions (cf. Def.~\ref{def:data:conditions}) on $V^\E$ to refer to the events' payloads.
%In addition, we use for every event in $e\in \E$ a dedicated variable $t_e$ of type integer to refer to the timestamp of $e$, and write $V_{time}(\E)$ for the set of all timestamp variables.
In the sequel we also assume that $V$ contains a special variable $\tmp$ of type integer, and $\tmp^\iota$ will denote the timestamp of event with id $\iota$.
To reason about partial orderings of events in $\E$, states use \emph{ordering conditions}, defined next:

\begin{definition}
An \emph{ordering condition} $o$ for a set of events $\E$ is of the following form, where $e,e'\,{\in}\,\E$, $a\,{\in}\,\Act$ is an activity, and $c$ is a data condition as in Def.~\ref{def:data:conditions}:
\[
o := e < e' \mid e \dirbefore e' \mid \mathit{first}(e) \mid \mathit{last}(e) \mid e \altbefore{a}{c} e' \mid \neg o \mid \top
\]
\end{definition}

Here $e < e'$ expresses that $e$ happens before $e'$, $e \dirbefore e'$ that
$e$ happens before $e'$ without any other event in between, $\mathit{first}(e)$ that $e$ is the first, $\mathit{last}(e)$ that $e$ is the last element, and $\top$ is a condition that is always true. Somewhat more complex, $e \altbefore{a}{c} e'$ expresses that $e$ happens before $e'$ without an event $e''$ in between that has activity $a$ and satisfies $c$, where $c$ is supposed to be a data condition over $V^{e''}$.
A set of ordering conditions on $\E$ in \emph{satisfiable} if there exists a topological sort of $\E$ that satisfies all conditions.
A set of ordering conditions $O$ is said to \emph{entail} an ordering condition $q$, denoted $O \models q$, if $\bigwedge O \wedge \neg q$ is unsatisfiable. 
Note that the ordering conditions are defined to closely align with the semantics of the supported Declare templates, as clarified in Def. \ref{def:corresp-order}. 

% For example, for the events
% $e_1=(\eid{1}, \m a, \{x=0\})$ and $e_2=(\eid{2}, \m b, \{x=2\})$, condition $e_1 < e_2$ expresses that $e_1$ happened before $e_2$, as in the trace of Ex.~\ref{exa:running:alignments}.

% We are now ready to give the definition of a state:

\begin{definition}
A \emph{state} is a pair $S =\tup{\E, C}$ where 
% \begin{compactitem}
% \item
$\E$ is a set of events, and
% \item 
$C$ is a set of ordering conditions on $\E$ and data conditions over $V^\E$.
% \end{compactitem}
\end{definition}

A state $\tup{\E, C}$ thus represents a set of events $\E$ that is partially ordered by the ordering conditions in $C$, and where payloads of events are constrained by the data conditions in $C$. 

For a trace $\trace=\tup{\seq[n]e}$, let $\E(\trace) = \{\seq[n]e\}$ be its set of events, $O(\trace) = \{e_i < e_{i+1} \mid 1 \leq i < n\}$ be the set of ordering conditions that capture the event ordering in $\trace$, and $D(\trace)$ the conjunction of all equations $v_\iota = \alpha(v)$ such that an event $e=(\iota,a,\alpha)$ occurs in $\trace$ and $v\in \dom(\alpha)$.
The \emph{initial state} is $\tup{\E(\trace), O(\trace) \cup D(\trace)}$, it serves as the starting point for the exploration of the search space.
Note that we use formulas that mix ordering and data conditions; we will explain in the next section how standard \smt{} solvers can be used to perform satisfiability checks of such formulas. 
% Note that using satisfiability checks, also entailments $\chi \models \xi$ can be checked, by ensuring that $\chi \wedge \neg \xi$ is unsatisfiable.

\begin{figure}[t]
\resizebox{\textwidth}{!}{
\includegraphics{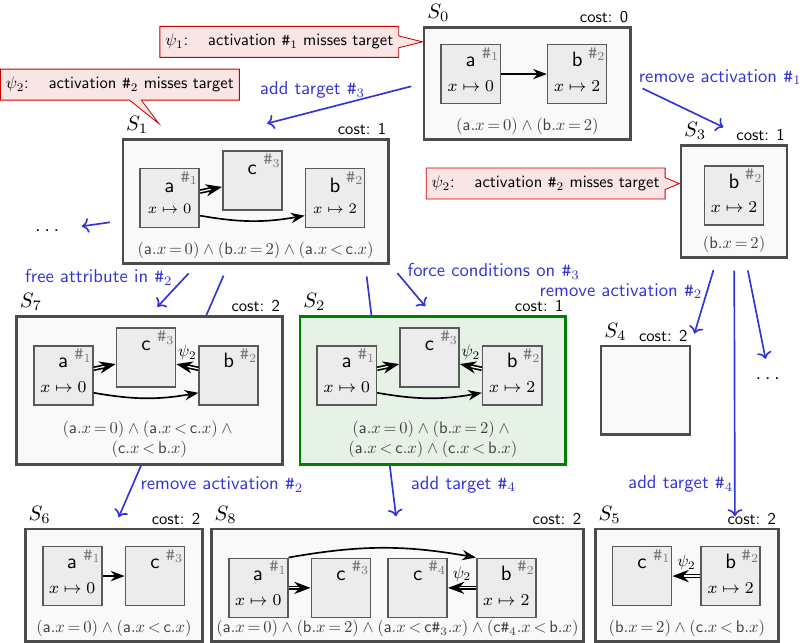}
}
\caption{Search space for the running example.\label{fig:search:space}}
\end{figure}

\begin{example}
Consider the trace $\trace$ in Ex.~\ref{exa:running:alignments} with events
$e_1=(\eid{1}, \m a, \{x=0\})$ and $e_2=(\eid{2}, \m b, \{x=2\})$.
 If no confusion can arise, we write $\m a.x$ rather than $x_{\eid{1}}$ etc. for readability.
The initial state is
$S_0=\tup{\{e_1,e_2\}, (e_1\,{<}\, e_2) \wedge (\m a.x\,{=}\,0) \wedge (\m b.x\,{=}\,2)}$.
Here $e_1\,{<}\, e_2$ expresses that $e_1$ happens before $e_2$; the remaining conditions fix the values of $x$ in the two events.
Fig.~\ref{fig:search:space} shows most of the search space for $\trace$ and the specification from Ex.~\ref{exa:running:spec} (the complete search space is shown in \reduced{\cite[Fig. 5]{extended}}\extended{Fig. \ref{fig:search:space:complete}}). States are shown as boxes, 
$S_0$ being the box on top.
The events of a state are shown as boxes with activities within the state, and arrows in between them indicate ordering conditions. Here $e_1 < e_2$ is displayed by an arrow 
\raisebox{-1mm}{\tikz{\node[inner sep=1pt](e1){$e_1$}; \node[inner sep=1pt,xshift=8mm](e2){$e_2$}; \draw[follows] (e1) -- (e2);}},
$e_1 \dirbefore e_2$ by 
\raisebox{-1mm}{\tikz{\node[inner sep=1pt](e1){$e_1$}; \node[inner sep=1pt,xshift=8mm](e2){$e_2$}; \draw[dirfollows] (e1) -- (e2);}}, and
the condition $e_1 \altbefore{\m b}{\m c.x < \m b.x} e_2$ obtained from $\psi_2$ by \raisebox{-1mm}{\smash{\tikz{\node[inner sep=1pt](e1){$e_1$}; \node[inner sep=1pt,xshift=9mm](e2){$e_2$}; \draw[dirfollows] (e1) -- node[above, scale=.5, near start] {$\psi_2$} (e2);}}}.
The formulas at the bottom of states specify data conditions.
The states $S_1$--$S_8$ are obtained from $S_0$ by applying repairs; we will explain below how this is done.
\end{example}

\subsection{Constraint violations}
We next define when constraints are violated in a state.
To that end, we need some additional notation:
given a \declare constraint $\psi$ and
events $e, e'\in \E$, we denote by $\Ord(\psi, e, e')$ the ordering conditions
imposed by $\psi$ between an activation event $e$ and a target event $e'$, defined as follows:

\begin{definition}\label{def:corresp-order}
Let $e,e'$ be events and $\psi=(\varphi, \cA, \cT, \cR)$ a constraint.
For templates $\varphi$ with an activation, we define $\Ord(\psi, e, e')$ as
$e<e'$  (resp.~$e'< e$) if $\varphi$ is based on a $\DC{Response}$ (resp.~$\DC{Precedence}$) template,
$e\dirbefore e'$ (resp. $e'\dirbefore e$) if it is a  $\DC{ChainResponse}$ (resp.~$\DC{ChainPrecedence}$) template,
% \sarahtodo{mini reformulation to avoid ugly spacing}
$\neg (e < e')$ for $\DC{NotResponse}$,
$\neg (e\dirbefore e')$ for $\DC{NotChainResponse}$, 
$e \altbefore{a}{\cA} e'$ for $\DC{AlternateResponse}$, and
$e' \altbefore{a}{\cA} e$ for $\DC{AlternatePrecedence}$.
In the last cases, $a$ is the activation activity of $\varphi$.
For constraints $\varphi$ without activation, let $\Ord(\psi, e)$ be $\mathit{first}(e)$ or $\mathit{last}(e)$ if $\varphi$ is an $\DC{Init}$ or $\DC{Last}$ constraint, respectively.
In all other cases, $\Ord(\psi, e)=\top$.
\end{definition}

We also need to instantiate data conditions for events. To that end,
given a \declare constraint $\psi=(\varphi, \cA, \cT, \cR)$ and an event $e=(\iota,a,\alpha)$ such
that $a$ is an activation activity for $\varphi$ and $b$ a target activity, 
we denote by $\varinst{\cA}{e}$ (resp. $\varinst{\cT}{e}$) the condition obtained from $\cA$ (resp. $\cT$) by substituting $v_a$ (resp. $v_b$) with $v_\iota$ for each $v\in V$.
Similarly, for another event $e'=(\delta,b,\alpha')$, $\varinst{\cT \wedge \cR}{e,e'}$ denotes the condition obtained from $\cT \wedge \cR$ by substituting variables $v_a$ by $v_\iota$, and $v_b$ by $v_{\delta}$ for all $v \in V$.

The first kind of violation is a \emph{missing target};
intuitively, it applies if a constraint $\psi$ can be activated but might lack a target that satisfies all conditions.

\begin{definition}
\label{def:missing:target}
A constraint $(\varphi, \cA{,}\cT{,}\cR) $ has a \emph{missing target} violation in state $(\E{,}C)$ if one of the following cases applies:
% \sarahtodo{added text to avoid a line with only 2 letters}
\begin{compactitem}
\item $\varphi$ is a response or precedence constraint with activation activity $a$ and there is an $e=(\iota,a,\alpha)\in \E$ such that 
$C \wedge \varinst{\cA}{e}$ is satisfiable,
but 
no $e'\in \E$ with target activity such that $\bigwedge C \wedge \varinst{\cA}{e} \models \Ord(\varphi, e, e') \wedge \varinst{\cT \wedge \cR}{e,e'}$; or
\item
$\varphi$ is of the form $\DC{Existence}(n,a)$, and $e_1,\dots,e_k$ are all events with activity $a$ in $\E$ but
$\bigwedge C \models \Sigma_{i=1}^k ite(\varinst{\cT}{e_i},1,0) \geq n$ does not hold; or
\item
$\varphi$ is an $\DC{Init}$, $\DC{End}$, or $\DC{Choice}$ constraint and $e_1,\dots,e_k$ are all events with target activity in $\E$ but
$\bigwedge C \models \bigvee_{i=1}^k\Ord(\psi, e_i) \wedge \varinst{\cT}{e_i}$ does not hold.
\end{compactitem}
\end{definition}

Here $ite(b,d_1,d_2)$ abbreviates an if-then-else expression.
In the first case of Def.~\ref{def:missing:target}, $e$ is called \emph{activation event}.

Fig.~\ref{fig:search:space} shows three cases of missing target violations for the constraints
in Ex.~\ref{exa:running:spec}:
in state $S_0$, $\psi_1$ is activated by the event $\eid{1}$ with activity $\m a$, but no target event with activity $\m c$ is present.
In $S_1$ and $S_3$, $\psi_2$ is violated: in $S_3$ since no event with activity $\m c$ occurs, and in $S_1$ because, even though an event with activity $\m c$ occurs, namely $\eid{3}$, its conditions do not entail $\m c.x < \m b.x$ and $\eid{3} \altbefore{a}{\cA} \eid{2}$.

The second kind of violation is dual in that it signals too many targets.

\begin{definition}
\label{def:excessive:target}
A constraint $\psi=(\varphi, \cA, \cT, \cR)$ has an \emph{excessive target} violation in a state $S=(\E,C)$ if one of the following cases applies:
\begin{compactitem}
\item $\varphi$ is of the form $\DC{Absence}(n,a)$ and there are $n$
events $e_1,\dots,e_n$ in $\E$ with activity $a$ such that
$\bigwedge C \wedge \bigwedge_{i=1}^n \varinst{\cT}{e_i}$ is satisfiable;
\item
$\varphi$ is a negation constraint, some $e_0,e_1\in \E$ have activation and target activity, resp., and
$\bigwedge C \wedge \Ord(\psi, e_0, e_1)\wedge
\varinst{\cA}{e_0} \wedge \varinst{\cT \wedge \cR}{e_0,e_1}$ is satisfiable.
\end{compactitem}
\end{definition}

The events $e_1,\dots, e_n$ in Def.~\ref{def:excessive:target} are called \emph{excessive target events}.

For instance, for the states in Fig.~\ref{fig:search:space}, a constraint $\tup{\DC{NotResponse}(\m a, \m b), \m a.x \geq 0, \top,\m b.x > \m a.x}$ would have an excessive target violation in states $S_0$ and $S_1$, but not in $S_3$. On the other hand, $\tup{\DC{Absence}(\m b), \top,\m b.x\,{=}\,3}$ would be violated in state $S_7$, but not in $S_0$ where the data conditions exclude $\m b.x\,{=}\,3$.

A constraint $\psi$ is \emph{violated} in a state if it has a missing or excessive target.
A state $S = \tup{\E, C}$ is a \emph{goal state} if no constraint in $\MM$ is violated in $S$ and $C$ is satisfiable.
In Fig.~\ref{fig:search:space}, all leaves of the search tree ($S_2$ and $S_4$--$S_8$) are goal states.

\subsection{Repairing violations}
Our approach subsequently expands the search space by selecting a state where a constraint is violated, generating \emph{child states} by repairing the violation in different ways.
Four kinds of repairs are distinguished:
\begin{inparaenum}[(a)]
\item addition of an event, which will be reflected as a model move in the alignment;
\item removal of an event that stems from the trace, corresponding to a log move in the alignment;
\item freeing a data attribute in an event that stems from the trace, corresponding to an edit move; and
\item enforcement of conditions. % corresponding to a synchronous move (not necessarily)
\end{inparaenum}

%However, 
The applicable repairs and resulting states depend on the violated constraint
$\psi = (\varphi, \cA, \cT, \cR) \in \MM$ and current state $S = \tup{\E,C}$.
First, if $\psi$ has an activation event $e_{\mathit{act}} \in \E$, the following
repairs are applied for both missing and excessive target violations to \emph{disable} the activation:
\begin{compactenum}[(1)]
\item 
\label{itm:repair:remove:act}
\textit{Removing an activation event.}
This repair only applies if $e_{\mathit{act}}$ stems from the trace $\trace$.
The resulting state is $S'=\tup{\E \setminus \{e_{\mathit{act}}\}, C'}$ where
$C'$ is like $C$ with conditions involving $e_{\mathit{act}}$ removed.
\item
\textit{Freeing a data attribute.}
\label{itm:repair:free:data:act}
This applies to an event $e_{\mathit{act}}=(\iota,a,\alpha)$ from the trace $\trace$ if $\alpha$ does not satisfy $\neg \varinst{\cA}{e_{\mathit{act}}}$.
The repair removes an assignment $\alpha(v)$ of $e_{\mathit{act}}$ for some $v\in V$. For $C' = C \setminus \{v_\iota=\alpha(v)\} \cup \{ \neg \varinst{\cA}{e_{\mathit{act}}}\}$, the new state is $S' = \tup{\E,C'}$.
\item 
\textit{Enforcing the negated activation condition.}
\label{itm:repair:enforce:neg:act}
The resulting state is $S'=\tup{\E, C'}$ with $C' = C \cup \{\neg\varinst{\cA}{e_{\mathit{act}}}\}$.
\end{compactenum}

\noindent
If the violation is a missing target for $\psi$, then a target event can be added, or an existing event with the correct activity can be enforced to satisfy the data conditions, or in some cases events can be removed that block ordering conditions.
More precisely, the following repairs apply:
\begin{compactenum}[(1)]
\setcounter{enumi}{3}
\item 
\label{itm:repair:add:target}
\emph{Adding a target event.}
A new state is of the form $S'=\tup{\E \cup \{e\}, C'}$ where
$e=(\iota, a,\emptyset)$ is a new event with fresh identifier $\iota$.
If $\psi$ has an activation and $e'$ is the activation event, then
$C' = C \cup \{\Ord(\psi, e', e), \varinst{\cA}{e'}, \varinst{\cT \wedge \cR}{ e',e}\}$. 
Otherwise, 
$C'= C \cup \{\Ord(\psi, e), \varinst{\cT}{e}\}$. 
\item 
\textit{Freeing a data attribute.}
\label{itm:repair:free:data:target}
This applies to events $e = (\iota,a,\alpha)$ in $\E$ that stem from the trace $\trace$ and have the target activity but do not satisfy $\varinst{\cT \wedge \cR}{e',e}$ if $\psi$ has an activation event $e'$, resp. $\varinst{\cT}{e}$ otherwise.
The repair removes some assignment $\alpha(v)$ for $v\in V$, which can avoid the violation.
The new state is $S'=\tup{\E', C \setminus \{v_\iota=\alpha(v)\}}$ where
$\E'$ is like $\E$ but where
$e$ is modifed to $(\iota{,}a{,}\alpha')$ such that $\alpha'$ is like $\alpha$ except being undefined for $v$.
\item 
\textit{Enforcing conditions.}
\label{itm:repair:enforce:conditions}
This applies to an event $e\in \E$ with activity $a$, i.e., a potential target event.
The new state is $S=\tup{\E,C'}$, where
%Sarah dropped, because:
% * C \wedge \neg(...) is always satisfiable because otherwise there is no violation
% * if C \wedge ... is not satisfiable, an unsat state is created, which is no problem
%where both $\bigwedge C \wedge \Ord(\psi, e) \wedge \varinst{\cT\wedge \cR}{e',e}$ and $\bigwedge C \wedge \neg (\Ord(\psi, e) \wedge \varinst{\cT\wedge \cR}{e',e})$ are satisfiable if there is an activation event $e'$,
%resp. both  $\bigwedge C \wedge \Ord(\psi, e) \wedge \varinst{\cT}{e}$ and $\bigwedge C \wedge \neg (\Ord(\psi, e) \wedge \varinst{\cT}{e})$ are satisfiable otherwise.
if $\psi$ has an activation, and $e'$ is the activation event that caused the missing target, then
$C' = C \cup \{\Ord(\psi, e', e)\} \cup \{\varinst{\cA}{e'}, \varinst{\cT \wedge \cR}{e',e}\}$; otherwise, $C'=C \cup \{\Ord(\psi, e), \varinst{\cT}{e}\}$.
\item 
\textit{Removing a blocking event.}
\label{itm:repair:remove:block}
This applies if an event $e_t\in \E$ with activity $a$ is according to the ordering conditions in $C$ not in the right position to act as target for $\psi$, but removing another event $e$ from the trace can make room for $e_t$. E.g., $\DC{Init}$ constraints delete the first event, 
%$\DC{End}$ constraints remove the last event, 
and $\DC{ChainResponse}$ %(resp. $\DC{ChainPrecedence}$) 
constraints remove the events directly succeeding %(resp. preceding) 
the activation. % according to $C$.
The resulting state is $S'=\tup{\E \setminus \{e\}, C'}$ where
$C'$ is like $C$ with conditions on $e$ removed.
\end{compactenum}

\noindent
If $\psi$ has an excessive target, we can either remove an excessive target event, or change the data conditions such that an event with target activity no longer acts as a target. 
Precisely, the following fixes apply:
\begin{compactenum}[(1)]
\setcounter{enumi}{7}
\item 
\emph{Removing excessive target events.}
\label{itm:repair:remove:extra:target}
This works like (\ref{itm:repair:remove:act}) above, but removes excessive target events if they stem from the trace.
\item 
\textit{Freeing a data attribute.}
\label{itm:repair:free:extra:target}
This repair is similar to (\ref{itm:repair:free:data:act}) above, but it applies if there is an excessive target event $e = (\iota,a,\alpha)$ in $\E$ that stems from the input trace.
However, we now enforce the negation of target and correlation conditions.
The resulting state is $S'=\tup{\E', C'}$ where
$\E'$ is like $\E$ but where
$e$ is modified to $(\iota,a, \alpha')$ such that $\alpha'$ is like $\alpha$ except that it is undefined for $v \in V$.
Let $\widehat C = C \setminus \{v_\iota=\alpha(v)\}$. 
If there is an activation event $e'$, we set 
$C' = \widehat C \cup \{ \neg\varinst{\cT \wedge \cR}{e',e}\}$; otherwise, $C' = \widehat C \cup \{\neg\varinst{\cT}{e}\}$.
\item 
\textit{Enforcing negated conditions.}
\label{itm:repair:enforce:neg:extra:target}
Let $e\in \E$ be an excessive target event.
There are two resulting states $S'=\tup{\E, C'}$ and $S''=\tup{\E, C''}$.
If there is an activation event $e'$, then
$C' = C \cup \{ \neg\varinst{\cT \wedge \cR}{e',e}\}$; otherwise, $C' = C \cup \{\neg\varinst{\cT}{e}\}$. Moreover,
$C'' = C \cup \{\neg \Ord(e',e)\}$.
\end{compactenum}
Note that \emph{all} applicable repairs are applied in all possible ways. For instance, when freeing a data attribute, a new state is generated for every event $e$ and every variable assignment in $e$ that satisfies the conditions in (\ref{itm:repair:free:data:act}). Also, if there is a missing target violation and $\varphi$ is a $\DC{Choice}$ constraint having two targets, a child state is created for each possible target. %\todo{Ale: is there some application strategy for the repairs? For example, the order/priority of application. If not, would a strategy of this type benefit the procedure? S: no, all repairs applied}

For example, in Fig.~\ref{fig:search:space}, 
$S_1$ is obtained from $S_0$ by adding a target event $\eid{3}$ (repair (\ref{itm:repair:add:target}));
$S_3$ is obtained from $S_0$ by removing the activation event $\eid{1}$ (repair (\ref{itm:repair:remove:act}));
$S_7$ is obtained from $S_1$ by freeing the data attribute $x$ in event $\eid{2}$ (repair (\ref{itm:repair:free:data:act})); and
$S_2$ is obtained from $S_1$ by forcing conditions on event $\eid{3}$ (repair (\ref{itm:repair:enforce:conditions})).

\subsection{A$^*$-based search}
Starting from the initial state that represents the input trace $\trace$, our algorithm subsequently chooses a state with a violation and generates child states by applying all possible repairs.
% From each goal state $S_g$, an alignment $\gamma(S_g)$ can be extracted, as we will explain below.
By a \emph{search space} for $\trace$ and $\MM$, we mean below a graph of states where the root is $S_0$, and all states have as children the states obtained by all possible repairs, if any.
To guide the search, the A$^*$ algorithm maintains for each state $S$  a cost $\cost(S) \in \mathbb R$, which can be shown to match exactly the cost of alignments extracted from $S$.
The initial state has cost 0.
When expanding a state $S$, the cost of a child state $S'$ is determined by the applied repair: when adding or removing events, or freeing a data attribute, we have $cost(S')=cost(S)+1$; when forcing condition satisfaction, $cost(S')=cost(S)$. %+\epsilon$.
In Fig.~\ref{fig:search:space}, each state is labelled with its respective cost.
% Our correctness results below show that the cost of a goal state $S_g$ is the cost of its alignment $\gamma(S_g)$. 

As repairs cause an increase in cost, by a fair exploration of the search space, the A$^*$ algorithm can conclude at some point that all goal states possibly detected in the future, will have a higher or equal cost than the goal states found so far.
At this point the search terminates, returning a goal state with minimal cost.

\newcommand{\modelEvents}{\mathit{modelEvents}}
\newcommand{\movelist}{\mathit{moves}}
\newcommand{\append}{\mathit{append}}
\newcommand{\modevent}{\mathit{modelEvent}}
\newcommand{\logevent}{\mathit{logEvent}}
\newcommand{\orIndex}{\mathit{idx}}
\newcommand{\lIndex}{{\mathit{i}}}
\newcommand{\mIndex}{{\mathit{j}}}
\newcommand{\model}{\mathit{model}}
\subsection{Alignment extraction}
From a goal state $S = \tup{\E, C}$, we extract an alignment
as described by the pseudocode in \Cref{alg:extract-alignment}.
The first step is to obtain an \smt{} model $\mu$ of the conditions $\bigwedge C$.
This induces a list of model events $\mathbf{f} = \tup{f_0, \dots, f_{m-1}}$ that satisfies all ordering conditions, and where for each $f_j = (\iota_j, a_j, \alpha_j)$ the assignment $\alpha_j$ is given by $\alpha_j(v) = \mu(v_{\iota_j})$ for all $0 \leq j < m$.

\Cref{alg:extract-alignment} then walks simultaneously along $\trace$ and $\mathbf{f}$, using $i$ as an index for $\trace$ and $j$ for $\mathbf{f}$, and adds an edit or synchronous move if the current events $e_i$ and $f_j$ share the same id (so $f_j$ stems from a trace event $e_i$), a model move if the id of the model event $f_j$ does not occur in $\trace$, and otherwise a log move.
(For an event $e=(\iota,a,\alpha)$, we write $e.id$ to refer to $\iota$.)
In Line~\ref{alg:extract-optimal-alignment:add-sync-move}, the number of mismatching assignments in $e_i$ and $f_j$ determines whether the move is an edit or synchronous move.
Note that the alignment extracted from a state is in general not unique as there can be multiple \smt{} models.
%The algorithm also applies to non-goal states, but the obtained alignment need not satisfy all constraints. % Not relevant
For instance, by applying \Cref{alg:extract-alignment} to state $S_2$ resp. state $S_6$ in Fig.~\ref{fig:search:space}, one obtains the alignments $\gamma_1$ resp. $\gamma_2$ shown in Ex.~\ref{exa:running:alignments}.

\begin{algorithm}
\caption{Extracting an alignment from a state}
\label{alg:extract-alignment}
\begin{algorithmic}[1]
\Require{State $S=\tup{\E, C}$,  trace $\trace=\tup{e_0,\dots, e_{n-1}}$}
\Ensure{Alignment for $\trace$}
\State $\model \leftarrow$ \smt{} model of formula $\bigwedge C$ \label{alg:extract-optimal-alignment:get-model}
\State $\tup{f_0, \dots, f_{m-1}} \leftarrow$ sort events in $\E$ by assignment to ordering conditions in $\model$ \label{alg:extract-optimal-alignment:sort-events}
\State $\movelist \leftarrow []$, $\lIndex \leftarrow 0$, $\mIndex \leftarrow 0$ \label{alg:extract-optimal-alignment:init-index}
\While{$(\lIndex < n) \vee (\mIndex < m)$} \label{alg:extract-optimal-alignment:loop}
    \If{$(\lIndex < n) \wedge (\mIndex < m) \wedge (e_\lIndex.id = f_\mIndex.id)$} \label{alg:extract-optimal-alignment:check-trace} 
        \State $\movelist.\append(\mathit{editOrSynchronousMove}(e_\lIndex, f_\mIndex))$ \label{alg:extract-optimal-alignment:add-sync-move}
        \State $\lIndex \leftarrow \lIndex + 1$, $\mIndex \leftarrow \mIndex + 1$
    \ElsIf{$(\mIndex < m)$ and $f_\mIndex.id$ does not occur in $\trace$}
        \State $\movelist.\append(\mathit{modelMove}(f_\mIndex))$ \label{alg:extract-optimal-alignment:add-model-move}
        \State $\mIndex \leftarrow \mIndex + 1$
    \Else
        \State $\movelist.\append(\mathit{logMove}(e_{\lIndex}))$ \label{alg:extract-optimal-alignment:add-log-move}
        \State $\lIndex \leftarrow \lIndex + 1$
    \EndIf
\EndWhile
\State \Return $\movelist$ \label{alg:extract-optimal-alignment:return-alignment}
\end{algorithmic}
\end{algorithm}

Our correctness result below shows that the alignment extracted from $S$ is optimal with cost $K$
(cf. the proof in~\reduced{the extended version~\cite[Sec. A.2]{extended}\ignorespaces}\extended{Sec. \ref{sec:proofs}\ignorespaces}).
Our running example illustrates this result: in Fig.~\ref{fig:search:space}, the goal state with mini\-mal cost is $S_2$, and indeed the optimal alignment $\gamma_1$ is extracted from it (cf. Ex.\,\ref{exa:running:alignments}).
%\todo{Ale: the A*-part of the algorithm is not decribed in Algorithm 1, right? So, we leave it only described 'by words'? If so, I would make the statement of Theorem 1 a bit more precies...}
%\sarahtodo{in which sense?}

% \begin{restatable}{lem}{alignments}
% If $S$ is a goal state then $\gamma$ returned by \Cref{alg:extract-alignment} on input $S$ and $\trace$ is an alignment of $\trace$ wrt. $\MM$.
% \sarahtodo{this lemma can also be dropped}
% \end{restatable}

\begin{restatable}[Correctness]{thm}{correctness}
If $S$ is a goal state with minimal cost $K$ in a search space for $\MM$ and $\trace$ then the list of moves $\gamma$ returned by \Cref{alg:extract-alignment} on input $S$ and $\trace$ is an optimal alignment of $\trace$ wrt. $\MM$ with cost $K$.
\end{restatable}

\section{Implementation}
\label{sec:implementation}

Our approach has been implemented in the tool {\toolname} written in Kotlin, using the \smt{} solvers Z3~\cite{Moura2008} and Yices~\cite{Dutertre2014} as backends.
{\toolname} requires two inputs: a multi-perspective event log in XES format~\cite{XES2023} and a Declare specification with data $\MM$. The model format is backward compatible with the one %in the datasets 
of~\cite{Bergami2021}. Nevertheless, the syntax for data conditions has been significantly enhanced, allowing users to leverage the full expressiveness of the \smtlib{} language~\cite{Barrett2025}. 
% Specifically, the \verb|content| of an \verb|(assert content)| expression can be utilized as a data condition or as a term within the simplified data condition syntax.
% \sarahtodo{what is \texttt{content} and \texttt{assert content}?\\Jacobo: The \smtlib{} language has assert expressions, and those are the ones supported (as they are required to return a boolean value). Could be written as any \smtlib{} boolean formula. Sarah: ok, I remember. but I still don't understand the sentence. if it's not super important, perhaps we can drop it to save space ...}
Also, the cost function can be customized, providing the cost of log, model, and edit moves as inputs. 
The \declare{} constraints language supported by our approach is, in fact, more expressive than initially introduced in Sec.~\ref{sec:preliminaries}. Specifically, branching in \declare{} constraints is enabled, as described in~\cite{CasasRamos2025}, and all \declare{} templates listed in ~\cite[Tab.~2]{CasasRamos2025} have been implemented.  
The tool produces an optimal alignment in a human-readable format, similar to Ex.~\ref{exa:running:alignments}. It can also export the search space as a graph, like in Fig.~\ref{fig:search:space}.
\smallskip

\noindent
\emph{Encoding.}
The \smt{} solver reasons on control flow and data dependencies in tandem, to identify violations and possible repairs for each constraint. We thus need to check satisfiability of formulas that mix ordering and data conditions.
Data conditions as in Def.~\ref{def:data:conditions}, but also much richer conditions, can be directly expressed in \smtlib{}. 
Ordering conditions on a set $\E$ are encoded as follows: for every event $e\in \E$ we use the \smt{} variable $\tmp_e$ of integer type that encodes the event's timestamp.
Then an ordering constraint $e_1 < e_2$ is directly translated to $\tmp_{e_1} < \tmp_{e_2}$;
$\mathit{first}(e)$ is translated to $\bigwedge_{e' \in \E \setminus\{e\}} \tmp_e < \tmp_{e'}$ and similar for $\mathit{last}(e)$; and
$e \dirbefore e'$ is translated to $\tmp_{e_1} < \tmp_{e_2} \wedge \bigwedge_{e \in \E \setminus\{e_1,e_2\}} (\tmp_{e} > \tmp_{e_2} \vee \tmp_{e} < \tmp_{e_1})$.
A constraint $e_1 \altbefore{a}{c} e_2$ is translated to $\tmp_{e_1} < \tmp_{e_2} \wedge \bigwedge_{e \in \E_a} (\neg \varinst{c}{e} \vee \tmp_{e} > \tmp_{e_2} \vee \tmp_{e} < \tmp_{e_1})$, where 
$\E_a$ is the set of all events in $\E$ with activity $a$, with $e_1$ and $e_2$ excluded.
Moreover, for efficiency, we use the SMT solver's assumption mechanism to temporarily check conditions, such as those in Defs.~\ref{def:missing:target} and \ref{def:excessive:target}. This approach allows us to assert temporary assumptions on top of a core set of formulas, avoiding unnecessary re-computations and improving performance.

\smallskip
\noindent
\emph{Optimizations.} We mention the most influential optimizations. 
The first is \textit{selecting a violation to repair:}
%before detecting violations, events that meet the activation condition are identified for each constraint and state, and kept during search.
as violations can be processed independently, the implementation selects for repair the one that generates the fewest child states, to delay the state explosion. 
%helps avoid state explosion. 
%choice of the violation to repair next is crucial
%, with profound effects on 
% for
% performance. 
% Currently, the tool selects
%The current approach involves selecting 
% the violation resulting in the 
%least number of 
% fewest child states, to
%criterion provides a reasonable starting point for the algorithm, delaying 
% delay the state explosion. 
%of the search space by creating very few branches toward the start of the search. 
% However, other selection strategies could improve performance for some cases, in particular in cases where the search space is highly complex or contains many interdependent conditions. Exploring other approaches is left as future work.

The second optimization is about \textit{detecting dead-ends:}
in every state, all violations are precomputed, and for every violation it is checked
which repairs are applicable. In case no repair is applicable for some violation, the 
state is a dead end, and the branch can be pruned from the search space.
% After identifying violations in the current state, the next step is to determine which violation to address first. 
% It is worth noting that all violated activations are precomputed, even if only one is selected for immediate repair. This allows for early detection and pruning of dead-ends. Dead-ends occur when no violation of a state can be repaired, despite the presence of other violated activations that can be repaired. In these cases, we know that there is no viable way to repair one of the errors, so it does not make sense to keep trying to repair other activations since it will not be possible to repair all of them. Hence the state can be removed from the search, pruning a whole branch of descendants from the search space. 
% For example, this can happen when the only way to fix a violation is to enforce the target event to satisfy the data constraint of $x \geq 0$, but that same data attribute  $x$ was previously enforced to be less than $0$ as a fix for another constraint. 

Finally, we \textit{prune unsatisfiable states.}
If a state $S=\tup{E,C}$ was generated where $\bigwedge C$
is unsatisfiable, conflicting conditions were added while generating the state. Therefore, the state can be dropped from the search space.

\section{Evaluation}
\label{sec:evaluation}

%We conduct our evaluation using the same Java Virtual Machine\footnote{OpenJDK 64-Bit VM Temurin 21.0.6+7-LTS} to execute all of the tested algorithms. The experiments are run on an Intel 5220R CPU with 8 GB of RAM. 
In our evaluation we execute all tested algorithms in the same environment, namely a Java Virtual Machine\footnote{OpenJDK 64-Bit VM Temurin 21.0.6+7-LTS} run on an Intel 5220R CPU with 8 GB of RAM.
The source code, executable, dataset and raw results are publicly available.\footnote{\url{https://apps.citius.gal/dada} and \url{https://doi.org/10.5281/zenodo.15470077}}

\smallskip
\noindent
\emph{Dataset.}
The evaluation utilizes a synthetic dataset that systematically varies in complexity, originally introduced in~\cite{Bergami2021}.
The complexity of the process models is influenced by the number of constraints (3, 5, 7, or 10) and constraint modifications (replacing 0, 1, 2, or 3 constraints). For each model, multiple event logs with varying trace lengths were generated (10, 15, 20, 25, or 30 events), resulting in 68,000 trace-model pairs.
The models feature simple variable-to-constant conditions over \texttt{categorical} (with values c1, c2, or c3) and \texttt{integer} (ranging from 0 to 100), such as \texttt{categorical is c1} or \texttt{integer > 10}.

\begin{figure}[t]
  \centering
  \includegraphics[width=0.85\textwidth]{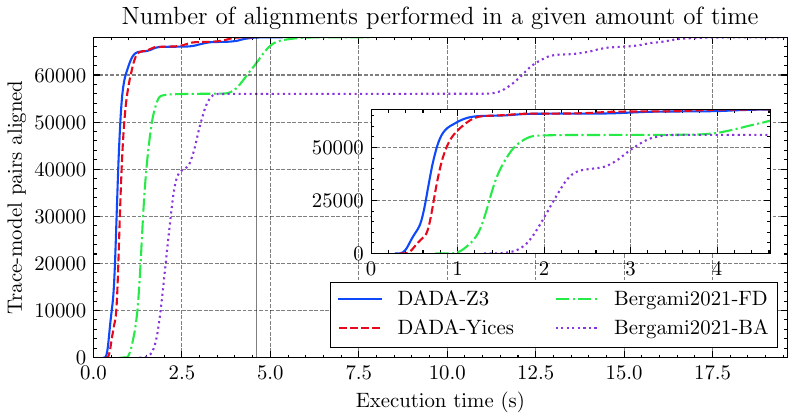}
  \caption{Number of trace-model pairs aligned within a given time frame by each algorithm. The enlargement %zoomed-in version % Sarah save a line
  on the right highlights the differences for shorter time intervals.}\label{fig:staircase}
\end{figure}

\smallskip
\noindent
\emph{Performance comparison.}
We compare DADA, using either the Z3~\cite{Moura2008} \smt{} solver or the Yices~\cite{Dutertre2014} \smt{} solver, to Bergami2021 \cite{Bergami2021}, using the original SymBA*~\cite{Torralba2014} planner or the Fast Downward~\cite{Helmert2006} planner. Our experiments measure the execution time for each pair of model and trace.
To ensure all alignments are optimal, we validate that the alignment costs produced by DADA-Z3 and DADA-Yices match those generated by Bergami2021-BA and Bergami2021-FD.

Fig.~\ref{fig:staircase} shows how, as the complexity of the trace-model pairs increases, the state-of-the-art algorithms exhibit a sharp increase in execution times, whereas our approach demonstrates better scalability.
Notably, our approach aligns any trace-model pair in at most 5 seconds, and DADA-Z3 is on average 2.9 times faster than Bergami2021-FD and 5.9 times faster than Bergami2021-BA

\begin{figure}[t]
  \centering
  \includegraphics[width=0.85\textwidth]{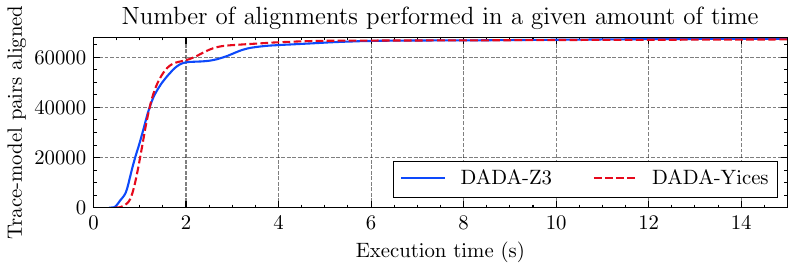}
  \caption{Performance evaluation incorporating correlation constraints.}\label{fig:staircase-correlation}
\end{figure}

\smallskip
\noindent
\emph{Constraint flexibility.}
While the previous experiment was limited to the data conditions supported by \cite{Bergami2021}, our approach can leverage the power of \smt{} solvers to define complex data dependencies such as the following correlation conditions.
In these conditions, \texttt{A} refers to the activation and \texttt{T} to the target; \texttt{cat} is an abbreviation for the \texttt{categorical} attribute, and \texttt{timestamp} is the event's time.

\begin{enumerate}[label=(C\arabic*), leftmargin=25pt, nolistsep]
    \item \texttt{A.timestamp + A.integer * 1d > T.timestamp + T.integer * 1d}
    \item \texttt{(A.cat - "0") \% 10 < (T.cat - "0") \% 10}
    \item \texttt{(A.cat == T.cat) ? (T.cat \% 2 == 0) : (A.integer > T.integer)}
\end{enumerate}

We create a new dataset by adding random negations, disjunctions and conjunctions of the previous correlation conditions to the original constraints, while retaining the original activation and target conditions, resulting in models like:

\noindent
\verb/Response[activity 1, activity 2]|...|...|¬(¬C1 or ¬(¬C3 and ¬C2))|/ 

\noindent
\verb/Chain Response[activity 3, activity 4] |...|...|¬C1 or ¬C2 or C3|/ 
% Sarah removed
%\verb/Absence[act. 5] |...|/ 

% A trace that previously was perfectly fitting the model ---and therefore had an optimal alignment of cost 0--- now requires a couple of edit moves, one for each of the modified constraints of the model. In this case, both edit moves are related to (C1), but each of them chooses a different variable to edit: the integer of the activation and the timestamp of the target respectively.

% \begin{verbatim}
% Alignment of cost 2:
%   | ... | act. 1{integer=29}  | ... | act. 4{time=2021-03-15...}             
%   | ... | act. 1{integer=100} | ... | act. 4{time=2021-03-18...}
% \end{verbatim}
% \sarahtodo{can we explain in one sentence what is \texttt{act.}i, and what is $\mathtt{C}_i$?}
The added correlation conditions make the alignment problem even harder by potentially increasing
\begin {enumerate*}[(a) ]%
\item the number of repairs required to reach the optimal alignment,
\item the number of ways in which it is possible to repair them, and
\item the work performed by the \smt{} solver within each state.
\end {enumerate*}
For these reasons, the models were simplified by only considering the ceiling of half of 
% their constraints.
the constraints generated in this way.
% \sarahtodo{slight reformulation to make more understandable (?)}
Fig.~\ref{fig:staircase-correlation} shows that our approach can handle these advanced conditions, with only a small percentage of alignments timing out or running out of memory (0.06\% for DADA-Yices and 0.91\% for DADA-Z3).

\section{Conclusions}
\label{sec:conclusions}

This paper presents a novel approach to computing data-aware optimal alignments between event logs and declarative process models, combining A* search and \smt{} solvers. Our key contributions include a new encoding scheme for the control flow, using an \smt{} solver to reason about control flow and data conditions, and an efficient A*-based search strategy that resolves constraint violations through repair actions.
Notably, the available constraint language is much richer than in earlier work, including a wide range of constructs supported by current SMT solvers.
We prove its correctness and demonstrate its efficiency in experiments, matching or surpassing state-of-the-art performance while supporting more expressive data dependencies. Future work includes exploring further optimizations such as advanced pruning strategies and heuristic functions.

\subsubsection{Acknowledgments.}
This work was partially funded by the Spanish Ministerio de Ciencia [grant numbers PID2020-112623GB-I00, PID2023-149549NB-I00, TED2021-130374B-C21], co-funded by the European Regional Development Fund (ERDF). J. Casas-Ramos gratefully acknowledges the support of CiTIUS for funding his research stay.
M.~Montali was partially supported by the NextGenerationEU FAIR PE0000013 project MAIPM (CUP C63C22000770006) and the PRIN MIUR project PINPOINT Prot. 2020FNEB27.
A. Gianola was partly supported by Portuguese national funds through Fundação para a Ciência e a Tecnologia, I.P. (FCT), under project UIDB/50021/2020 (DOI: 10.54499/UIDB/50021/2020). This work was partially supported by the `OptiGov' project, with ref. n. 2024.07385.IACDC (DOI: 10.54499/2024.07385.IACDC), fully funded by the `Plano de Recuperação e Resiliência' (PRR) under the investment `RE-C05-i08 - Ciência Mais Digital' (measure `RE-C05-i08.m04'), framed within the financing agreement signed between the `Estrutura de Missão Recuperar Portugal' (EMRP) and Fundação para a Ciência e a Tecnologia, I.P. (FCT) as an intermediary beneficiary.

\subsubsection{Disclosure of Interests.}
The authors have no competing interests to declare.

\bibliographystyle{splncs04}
\bibliography{references}

\extended{
\newpage
\appendix
\section{Appendix}
In this appendix, we first define formally the semantics of \declare constraints with data, and then provide a formal correctness proof of our approach.

\subsection{Semantics of \declare with Data}

The following definition clarifies when a trace satisfies \declare constraints with data.
For an assignment $\alpha$ with domain $V$, we write $\alpha^a$ for some $a\in \Act$ for the same assignment on labeled variables, i.e., the assignment with domain $\{v^a \mid v\in V\}$ that sets $\alpha^a(v^a)=\alpha(v)$.
Moreover, we write $\alpha \models c$ to express that $\alpha$ satisfies a condition $c$.
For the union of two assignments $\alpha,\beta$ with disjoint domain we write $\alpha \cup \beta$.

\begin{definition}\label{def:declare:constraints}
A constraint $\psi=\tup{\varphi, \cA, \cT, \cR}$ is satisfied by a trace $\trace=\tup{e_0,\dots,e_{m-1}}$ if
\begin{compactitem}
\item 
$\varphi=\DC{Existence}(n,a)$, there are $n$ distinct events $e_{i_1}, \dots, e_{i_n}$ in $\trace$ such that for all $1\leq j \leq n$ and if $e_{i_j}$ has the form $e_{i_j}=\tup{\iota_j,a,\alpha_j}$ it holds that $\alpha^a_j \models \cT$;
\item 
$\varphi=\DC{Absence}(n,a)$, and $\trace$ does not satisfy $\tup{\DC{Existence}(n,a), \cA, \cT, \cR}$;
\item 
$\varphi=\DC{Init}(a)$, $e_0=\tup{\iota,a,\alpha}$, and $\alpha^a \models \cT$;
\item 
$\varphi=\DC{End}(a)$, $e_{m-1}=\tup{\iota,a,\alpha}$, and $\alpha^a \models \cT$;
\item 
$\varphi=\DC{Choice}(a,b)$, and there is some $e_i=\tup{\iota,d,\alpha}$, $1\leq i < m$, such that $d=a$ and $\alpha^a \models \cT$, or $d=b$ and $\alpha^b \models \cT$;
\item 
$\varphi=\DC{RespondedExistence}(a,b)$, and either there is no $e_i=\tup{\iota,a,\alpha}$, $0\leq i < m$, such that $\alpha^a \models \cA$, or there is some $e_j=\tup{\iota',b,\beta}$, with $0\leq j < m$ and $i\neq j$, such that $\alpha^a \cup \beta^b \models \cT \wedge \cR$;
\item 
$\varphi=\DC{Response}(a,b)$, and either there is no $e_i=\tup{\iota,a,\alpha}$, $0\leq i < m$, such that $\alpha^a \models \cA$, or there is some $e_j=\tup{\iota',b,\beta}$, with $i < j < m$, such that $\alpha^a \cup \beta^b \models \cT \wedge \cR$;
\item 
$\varphi=\DC{AlternateResponse}(a,b)$, and either there is no $e_i=\tup{\iota,a,\alpha}$, $0\leq i < m$, such that $\alpha^a \models \cA$, or there is some $e_j=\tup{\iota',b,\beta}$, with $i < j < m$, such that $\alpha^a \cup \beta^b \models \cT \wedge \cR$, and for all $e_k$ with $i < k < j$ of the form $e_k=\tup{\iota', d, \alpha_k}$ either $d \neq a$ or $\alpha^a \not\models \cA$;
\item 
$\varphi=\DC{ChainResponse}(a,b)$, and either there is no $e_i=\tup{\iota,a,\alpha}$, $1\leq i \leq n$, such that $\alpha^a \models \cA$, or $i < m-1$ and $e_{i+1}=\tup{\iota',b,\beta}$ and $\alpha^a \cup \beta^b \models \cT \wedge \cR$;
\item 
$\varphi=\DC{Precedence}(a,b)$, and either there is no $e_i=\tup{\iota,b,\alpha}$, $0\leq i < m$, such that $\alpha^b \models \cA$, or $i > 0$ and there is some $e_j=\tup{\iota',a,\beta}$, with $0 \leq j < i$, such that $\alpha^b \cup \beta^a \models \cT \wedge \cR$;
\item 
$\varphi=\DC{AlternatePrecedence}(a,b)$, and either there is no $e_i=\tup{\iota,b,\alpha}$, $0\leq i < m$, such that $\alpha^b \models \cA$, or $i > 0$ and there is some $e_j=\tup{\iota',a,\beta}$, with $0 \leq j < i$, such that $\alpha^b \cup \beta^a \models \cT \wedge \cR$, and for all $e_k$ with $j < k < i$ of the form $e_k=\tup{\iota', d, \alpha_k}$ either $d \neq b$ or $\alpha^b \not\models \cA$;
\item 
$\varphi=\DC{ChainPrecedence}(a,b)$, and either there is no $e_i=\tup{\iota,b,\alpha}$, $0\leq i < m$, such that $\alpha^b \models \cA$, or $i > 0$ and $e_{i-1}=\tup{\iota',a,\beta}$ and $\alpha^b \cup \beta^a \models \cT \wedge \cR$;
\item 
$\varphi=\DC{NotResponse}(a,b)$, and $\trace$ does not satisfy $\tup{\DC{Response}(n,a), \cA, \cT,$ $\cR}$;
\item 
$\varphi=\DC{NotRespondedExistence}(a,b)$, and $\trace$ does not satisfy the constraint $\tup{\DC{RespondedExistence}(n,a), \cA, \cT, \cR}$; or
\item 
$\varphi=\DC{NotChainResponse}(a,b)$, and trace $\trace$ does not satisfy the constraint $\tup{\DC{ChainResponse}(n,a), \cA, \cT, \cR}$.
\end{compactitem}
\end{definition}

% \subsection{Running Example}
% The complete search space for the running example is shown in Fig.~\ref{fig:search:space:complete}.
\begin{figure}[ht]
\includegraphics[angle=90]{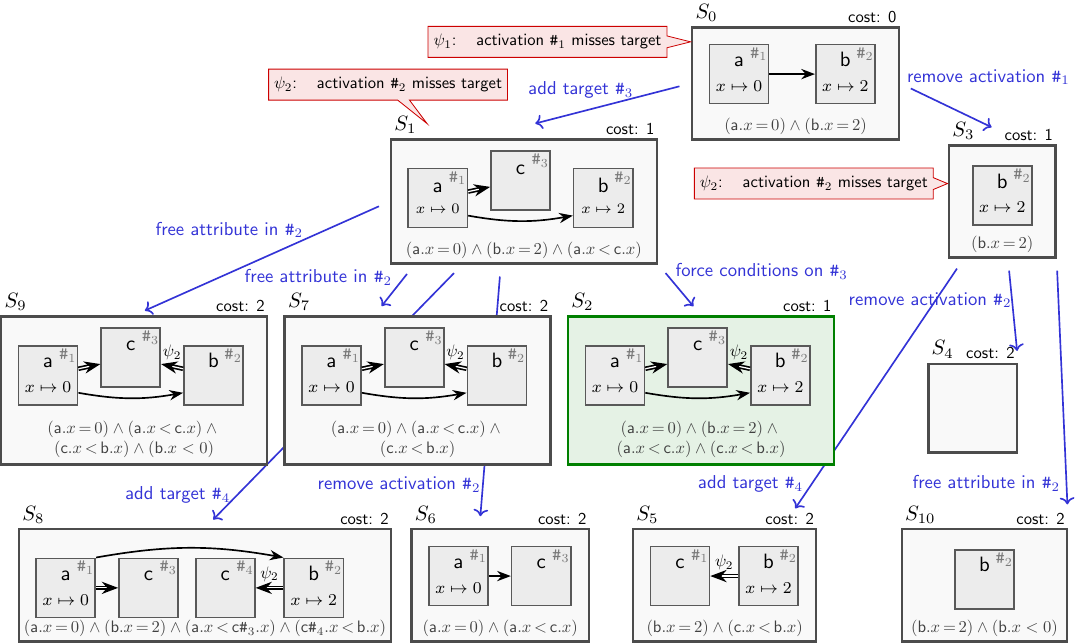}
\caption{Complete search space for running example\label{fig:search:space:complete}}
\end{figure}

\subsection{Correctness Proof}\label{sec:proofs}

\renewcommand{\phi}{\varphi}
%\sarahtodo[inline]{
%additional assumptions:\\
%removal is only applied to unmodified trace events\\
%when freeing data attribute, add its negation?
%}

The following is straightforward to show by a case distinction on $\psi$.

\begin{lemma}
\label{lem:alignment:model}
Let $\psi =(\varphi, \cA, \cT, \cR)$ be a constraint.
A state $S$ does not violate $\psi$ if and only if for all $\gamma \in \Gamma(S)$, it holds that $\restr{\gamma}{M}$ satisfies $\psi$.
\end{lemma}

We next show that if \Cref{alg:extract-alignment} returns  $\gamma$ on input $S$ and $\trace$ then $\gamma$ is indeed an alignment for $\trace$, though in general only wrt. a subset of $\MM$, and the cost of $\gamma$ coincides with $cost(S)$.
Below, for a fixed trace $\trace$, we denote by $\Gamma(S)$ the set of alignments that can be extracted from $S$, i.e., that are possible results of \Cref{alg:extract-alignment} on input $S$ (recall that the alignment extracted from a state $S$ is in general not unique).

Below, we make the following assumption $(\dagger)$ on the search space generated for $\MM$ and $\trace$: Repairs that remove an event $e$ are only applied if $e$ was never modified beforehand by a repair that changes the assignment.

\begin{lemma}[Soundness]
\label{lem:soundness}
For each $\gamma \in \Gamma(S)$, it holds that $\restr{\gamma}{L}=\trace$
% wrt. a subset of $\MM$ 
and $\kappa(\gamma) = cost(S)$. 
\end{lemma}
\begin{proof}
First of all, it is easy to see that if $S=\tup{\E,C}$ and $\gamma \in \Gamma(S)$ then $\restr{\gamma}{L}=\trace$ because \Cref{alg:extract-alignment} adds for every $e_i$ in $\trace$ a log or synchronous/edit move, since $i$ is incremented from $0$ to $n-1$.

It remains to show the claim about the cost of $\gamma$, which we prove 
by induction on the depth $n$ at which $S$ occurs in the search tree. The statement holds for the initial state $S_0$, where an alignment $\gamma \in \Gamma(S)$ consists of only synchronous moves, so $\kappa(\gamma)=0=cost(S_0)$.

Let $S'=\tup{\E',C'}$ be at depth $n+1$ in the search tree.
The induction hypothesis is that the claim holds for all states at level $n$.
We perform a case distinction on the repair applied at the parent $S=\tup{\E,C}$ of $S'$ to create $S'$.
\begin{compactenum}
\item 
If an event $e$ was added, then $e$ has a fresh id that does not occur in the trace.
Each alignment $\gamma' \in \Gamma(S')$ stems from some model $\mu$ for $C'$.
By construction of the repair (the constraints remain the same), $\mu$ also satisfies $C$, giving rise to an alignment $\gamma \in \Gamma(S)$. By the induction hypothesis, $\gamma$ is an alignment of $\trace$ with cost $cost(S)$.
Alignment $\gamma'$ must be like $\gamma$ except for an additional model move (as the added id is fresh, it cannot match an event in the trace), so that $\kappa(\gamma')=\kappa(\gamma)+1$.
Since we have $cost(S')=cost(S)+1$ and $\kappa(\gamma)=cost(S)$ by the induction hypothesis, the claim holds.
\item
If an event was removed, then this event stems from the trace, and was not modified beforehand by assumption ($\dagger$).
Each alignment $\gamma' \in \Gamma(S')$ stems from some assignment $\mu'$ for $C'$.
Ordering conditions and data conditions are independent.
Thus there is an assignment $\mu$ that coincides with $\mu'$ on ordering constraints and assigns arbitrary data values compatible with $C$.
Even though data values differ, for all moves except for the one concerning the removed event, their costs coincide, as required data values are enforced by dedicated conditions.
The corresponding alignment 
$\gamma\in \Gamma(S)$ has a synchronous move where $\gamma'$ has a log move, so $cost(\gamma')=cost(\gamma)+1$.
Since $cost(S')=cost(S)+1$ and $cost(\gamma)=cost(S)$ by induction hypothesis, the claim holds.
\item
Suppose a data attribute $v$ in an event $e=(\iota,a,\alpha)$ was freed. 
Each alignment $\gamma' \in \Gamma(S')$ stems from some assignment $\mu'$ for $C'$.
Ordering conditions and data conditions are independent.
Thus there is an assignment $\mu$ for $C$ that coincides with $\mu'$ on ordering constraints and assigns data values compatible with $C$, in particular $\mu(v^\iota)=\alpha(v)$.
Even though data values differ, for all moves except for the one concerning $e$, their costs coincide, as required data values are enforced by dedicated conditions.
Since $cost(S')=cost(S)+1$ and $cost(\gamma)=cost(S)$ by induction hypothesis, the claim holds.
\item
Suppose conditions were enforced.
Then any model of $C$ is also a model of $C'$ as $C' \subseteq C$, and 
each alignment $\gamma \in \Gamma(S)$ that stems from some model $\mu$ for $C$ is also an alignment in $\Gamma(S)$.
Since $cost(S)=cost(S')$, the claim follows from the induction hypothesis.
\qed
\end{compactenum}
\end{proof}

Let two alignments $\gamma$ and $\gamma'$ for $\trace$ and $\MM$ be \emph{equivalent} if
$|\gamma| = |\gamma'|$ and the move in $\gamma$ at position $i$ is a log/model/synchronous move if so is the move in $\gamma'$ at position $i$, and where edit moves edit the same variables. However, variable values in the model component of $\gamma$ and $\gamma'$ that do not match a variable in a trace may differ, as well as event identifiers.

\begin{lemma}[Completeness]
\label{lem:completeness}
Let $\gamma$ be an optimal alignment for $\trace$ and $\MM$.
Then there is a goal state $S_g$ in the search space such that $\Gamma(S_g)$ contains an alignment equivalent to $\gamma$.
\end{lemma}
\newcommand{\eact}{e_{\mathit{act}}}
\newcommand{\etgt}{e_{\mathit{tgt}}}
\begin{proof}
Let $\gamma$ be an optimal alignment for $\trace=\tup{e_1,\dots,e_n}$ and $\MM$.
Let $\gamma|_M=\run=\tup{f_0, \dots, f_m}$.

We show that there is a sequence of states $S_0, S_1, S_2, \dots, S_g $ in the search tree such that, intuitively, by descending along this path the respective alignment gets closer and closer to $\gamma$, and $S_g$ is a goal state that satisfies the claim.
More precisely, we show that for each state $S_i=\tup{\E,C}$, there is a correspondence relation $R_{S_i} \subseteq \E \times \{f_0, \dots, f_m\}$ associating some of its events with events in $\run$ that satisfies the following invariants:
\begin{compactenum}
\item[$(i)$] For every pair $(e,f)\in R_{S_i}$, the two events have the same activity.
\item[$(ii)$] For all events $e\in E \setminus \{e_1,\dots,e_n\}$, there is some $f_j$ in $\run$ such that $(e,f_j) \in R_{S_i}$. That is, all added events have a correspondent in $\run$. Moreover, for all edit or synchronous moves $(e,f_j)$ in $\gamma$, $f_j$ has a match in $R_{S_i}$. 
\item[$(iii)$] 
Let $E_R \subseteq E$ be the set of events  $\{e \in E \mid \exists f. (e,f)\in R_{S_i}\}$,
$\mu$ the (partial) assignment on $vars(C)$ that sets, for all $(e,f)\in R_S$ with $\iota$ the id of $e$, $\mu(\tau^\iota) = \mathit{time}(f)$ and $\mu(x^\iota)=\alpha(x)$, where $f=(\iota',a,\alpha)$.
Then $\mu$ satisfies $(C\setminus C_0)|_{V_R}$ where $V_R$ is the union of all $V^\iota$ such that $\iota$ is an id of an event in $E_R$.
%Let $\mu$ be the (partial) assignment on $vars(C)$ that sets the timestamps according to $\run$, i.e., $\mu(t_e)=time(f_j)$, and assigns payloads of $e$ like payloads in $f_j$, for all $(e,f_j) \in R_{S_i}$.
%Let $\E|_{R_{S_i}}$ be the subset of $\E$ given by $\E|_{R_{S_i}} = \{e \mid \exists f. (e,f)\in R_{S_i}\}$. 
%Then $\mu$ satisfies $C \setminus C_0$, where $C_0$ are the conditions from the initial state.
\item[$(iv)$] Only attributes of events are freed that have via $R_S$ a correspondent in $\run$.
\end{compactenum}
Moreover, the sequence of relations is monotonically increasing, i.e., we have $R_{S_0} \subseteq R_{S_1} \subseteq R_{S_2} \subseteq \dots$.

We first show existence of a sequence $S_0, S_1, S_2, \dots$  that satisfies the invariants, then we reason that it contains a goal state $S_g$ that satisfies the claim.

At depth $k=0$, we have $S=S_0$ and $\E=\{e_1,\dots, e_n\}$. 
Let $R_{S_0}$ consist of all pairs $(e_i,f_j)$ such that $(e_i,f_j)$ is a synchronous or edit move in $\gamma$.
The relation $R_{S_0}$ satisfies $(i)$ because in edit and synchronous moves the activity is shared. Item $(ii)$ and $(iii)$ are vacuously satisfied as no events were added, and $(iv)$ because no attributes were freed.

\smallskip
Consider now a state $S=\tup{\E,C}$ at depth $k$.
If $S$ is a goal state, we are done, so we assume that $S$ has a violation.
Suppose $S$ gets repaired for a constraint $\psi=\tup{\varphi, \cA,\cT,\cR}\in \MM$.
Consider first the case of a missing target violation; by a case distinction, we decide a next state $S'$.
\begin{compactenum}
\item
Suppose first that $\psi$ does not have an activation.
For simplicity, we consider the case of a single target, but the case for $\DC{Existence}$ is similar.
Let $f_j$ be the target of $\psi$ in $\run$.
\begin{compactenum}
\item
If there exists some $e=(\iota,a,\alpha)\in \E$ such that $(e,f_j) \in R$ then $e$ and $f_j$ must have the same activity by $(i)$, and we must have $C\not\models \Ord(\psi, e) \wedge \varinst{\cT}{e}$ ($\star$), otherwise there would be no violation.
Let $f_j=(\iota', a, \alpha')$.
\begin{compactitem}
\item
Suppose there is some $v\in \dom(\alpha)$ such that $\alpha(v) \neq \alpha'(v)$.
This is only possible if $e$ stems from the trace, since added events have no fixed values.
Thus, let $S'$ be the child state obtained from a repair (\ref{itm:repair:free:data:target}) where attribute $v$ was freed.
\item
Otherwise, if $\bigwedge C\wedge \Ord(\psi, e) \wedge \varinst{\cT}{e}$ is satisfiable,
let $S'$ be the result of a condition enforcement repair (\ref{itm:repair:enforce:conditions}).
The repair is applicable because as observed above $e$ and $f_j$ have the same activity, and the resulting state is satisfiable because $C\wedge \Ord(\psi, e) \wedge \varinst{\cT}{e}$ is satisfiable.
\item
Otherwise, $\bigwedge C \wedge \Ord(\psi, e) \wedge \varinst{\cT}{e}$ is not satisfiable.
Since $\mu$ satisfies $(C\setminus C_0)|_{V_R}$ and $\mu$ satisfies $\Ord(\psi, e) \wedge \varinst{\cT}{e}$, there must be some event $e'\in \E$ from the trace, i.e. which adds conditions to $C_0$, that causes unsatisfiability.
In fact, as assignments in $e$ and $f_j$ have no mismatches,
$\bigwedge C\wedge \Ord(\psi, e)$ must be unsatisfiable. Precisely, $\Ord(\psi, e)$ must be $\mathit{first}(e)$ (resp. $\mathit{last}(e)$), but $C$ contains $e' < e$ (resp. $e' > e$) for some trace event $e'$.
Then $e'$ cannot have a match in $R_S$ because $\mu$ satisfies $C\setminus C_0$ restricted to $E|_{R_S}$ by $(iii)$. Hence we can apply repair (\ref{itm:repair:remove:block}) to remove $e'$, obtaining a state $S'$.
Note that by invariant $(iv)$, no attribute in $e'$ can have been freed because it has no correspondent in $R_S$.
\end{compactitem}
In all these cases $R_S$ stays the same and thus satisfies conditions $(i)--(iii)$.
Moreover, in the first case the freed attribute belongs to an event that has a correspondent in $R_S$, so also $(iv)$ is satisfied.
\item
Now assume there is no match for $f_j$ in $R$.
Then $f_j$ cannot be in a synchronous or edit move in $\gamma$ by invariant $(ii)$.
So $f_j$ is in a model move, thus let $S'$ be the state obtained from a repair (\ref{itm:repair:add:target}) where a target event $e$ was added.
Set $R_{S'}=R_S \cup \{(e, f_j)\}$, which satisfies the invariants $(i)-(iii)$, and $(iv)$ is satisfied because no additional attribute is freed. 
\end{compactenum}
\item
Now suppose $\psi$ has an activation, let $\eact\in \E$ be the activation event of the violation.
\begin{compactenum}
\item
Suppose there is some $f_j$ such that $(\eact, f_j) \in R$.
% Then $f_j$ must be in a synchronous or edit move in $\gamma$.
Suppose first that for $f_j = \tup{\iota',a,\alpha'}$, $\alpha'$ does not satisfy $\cA$. 
\begin{compactitem}
\item
If there is some $v\in \dom(\alpha)$ such that $\alpha(v) \neq \alpha'(v)$ then
let $S'$ be the child state obtained from a repair (\ref{itm:repair:free:data:act}) where attribute $v$ was freed. 
\item
Otherwise, let $S'$ be the result of a condition enforcement repair (\ref{itm:repair:enforce:neg:act}).
\end{compactitem}

Second, suppose $\alpha'$ satisfies $\cA$, so it must have a target $f_k$ in $\run$, $k\neq j$.
\begin{compactenum}
\item
If there exists some $\etgt=(\nu,b,\beta)\in \E$ such that $(\etgt,f_k) \in R_S$ then we must have $C\not\models \Ord(\psi, \eact, \etgt) \wedge \varinst{\cT\wedge \cR}{\eact,\etgt}$ ($\star$), otherwise there would be no violation.
Let $f_k=(\nu', b, \beta')$.
\begin{compactitem}
\item
If there is some $v\in \dom(\beta)$ such that $\beta(v) \neq \beta'(v)$ then $\etgt$ must stem from the trace.
Let $S'$ be the child state obtained from a repair (\ref{itm:repair:free:data:target}) where attribute $v$ was freed.
\item
Otherwise, if $\bigwedge C\wedge \Ord(\psi, \eact, \etgt) \wedge \varinst{\cT\wedge \cR}{\eact,\etgt}$ is satisfiable, let $S'$ be the result of a condition enforcement repair (\ref{itm:repair:enforce:conditions}).
The repair is applicable because of ($\star$), and as $\bigwedge C\wedge \Ord(\psi, \eact, \etgt) \wedge \varinst{\cT\wedge \cR}{\eact,\etgt}$ is satisfiable, also the resulting state is satisfiable.
\item
Otherwise, if $\bigwedge C\wedge \Ord(\psi, \eact, \etgt) \wedge \varinst{\cT\wedge \cR}{\eact,\etgt}$ is unsatisfiable, this must be because of some trace events in $\E$ that have no correspondent in $\run$ via $R_S$, since $\bigwedge (C\setminus C_0)|_{V_R}\wedge \Ord(\psi, \eact, \etgt) \wedge \varinst{\cT\wedge \cR}{\eact,\etgt}$ is satisfied by $\mu$ according to property $(iii)$.
In fact, as assignments in $e$ and $f_j$ have no mismatches (this was already excluded above),
$\bigwedge C\wedge \Ord(\psi, \eact, \etgt)$ must be unsatisfiable. Precisely, $\Ord(\psi, \eact, \etgt)$ must be $\eact \dirbefore \etgt$ but $C$ contains $\eact < e'$ and $e' <  \etgt$ (or similar for precedence) for some trace event $e'$.
Then $e'$ cannot not have a match in $R_S$, and apply repair (\ref{itm:repair:remove:block}) to remove $e'$, obtaining a state $S'$.
Note that by $(iv)$ no attribute in $e'$ has ever been freed because $e'$ has no correspondent in $R_S$.
\end{compactitem}
\item
Now assume there is no match for $f_k$ in $R_S$.
Then $f_k$ cannot be in a synchronous or edit move in $\gamma$ by property $(ii)$.
So $f_k$ is in a model move. Let $S'$ be the state obtained from a repair (\ref{itm:repair:add:target}) where a target event $e$ was added.
Set $R_{S'}=R_S \cup \{(\etgt, f_k)\}$, which satisfies the invariants. 
\end{compactenum}
\item Suppose there is no $(\eact, f_j) \in R_S$. By invariant $(ii)$, $\eact$ stems from the trace. We then apply the repair (\ref{itm:repair:remove:act}) to remove the activation event $\eact$, obtaining a state $S'$. 
Note that the activation event cannot have been modified, otherwise there would be a match in $R_S$ by property $(iv)$.
\end{compactenum}
It can be checked that in all cases, the invariants $(i)-(iv)$ remain satisfied.
\end{compactenum}

\noindent
Second, consider an excessive target violation.
Let $\etgt$ be the excessive target event that caused the violation.
\begin{compactenum}
\setcounter{enumi}{2}
\item
Suppose first that $\psi$ does not have an activation.
For simplicity, we consider the case of a single target, but the case for $\DC{Absence}$ is similar.
\begin{compactitem}
\item
Suppose there is some $f_j$ such that $(\etgt, f_j) \in R_S$.
Suppose first that for $f_j = \tup{\iota',a,\alpha'}$, $\alpha'$ does not satisfy $\cT$. 
\begin{compactitem}
\item
If there is some $v\in \dom(\alpha)$ such that $\alpha(v) \neq \alpha'(v)$ then $\etgt$ must stem from the trace.
Let $S'$ be the child state obtained from a repair (\ref{itm:repair:free:extra:target}) where attribute $v$ was freed. 
\item
Otherwise, let $S'$ be the result of a condition enforcement repair (\ref{itm:repair:enforce:neg:extra:target}).
\end{compactitem}
It can be checked that in all cases, the invariants $(i)--(iv)$ remain satisfied.
\item
Suppose there is no $f_j$ such that $(\etgt, f_j) \in R_S$. Then $\etgt$ must stem from the trace by property $(ii)$. We apply repair (\ref{itm:repair:remove:extra:target}) to remove an extra target. 
\end{compactitem}
\item
Now suppose $\psi$ has an activation, let $\eact\in \E$ be the activation event of the violation.
\begin{compactenum}
\item
Suppose first there is some $f_j$ such that $(\eact, f_j) \in R_S$. 
\begin{compactenum}
\item Suppose there is some $f_k$ in $\run$ such that $(\etgt, f_k) \in R_S$.
Let $\etgt=(\nu,b,\beta)$ and $f_k=(\nu', b, \beta')$.
Since $\run$ satisfies $\psi$, the assignment $\mu$ cannot satisfy $\Ord(\psi, \eact, \etgt) \wedge \varinst{\cT\wedge \cR}{\eact,\etgt}$.
\begin{compactitem}
\item
If there is some $v\in \dom(\beta)$ such that $\beta(v) \neq \beta'(v)$ then $\etgt$ must stem from the trace.
Let $S'$ be the child state obtained from a repair (\ref{itm:repair:free:extra:target}) where attribute $v$ was freed.
\item
Otherwise, we apply a condition enforcement repair (\ref{itm:repair:enforce:neg:extra:target}).
If $\mu$ does not satisfy $\Ord(\psi, \eact, \etgt)$ then we take the state
$S' = \tup{\E,C'}$ where $C'=C\cup \{ \neg \Ord(\psi, \eact, \etgt) \}$.
Otherwise, $\mu$ does not satisfy $\varinst{\cT\wedge \cR}{\eact,\etgt}$, and we take the state $S'' = \tup{\E,C''}$ where $C'=C\cup \{ \neg \varinst{\cT\wedge \cR}{\eact,\etgt} \}$.
\end{compactitem}
\item Suppose there is no $f_k$ in $\run$ such that $(\etgt, f_k) \in R_S$. 
Then $\etgt$ must stem from the trace.
We apply repair (\ref{itm:repair:remove:extra:target}) to remove the target event.
\end{compactenum}
\item
Suppose there is no $f_j$ such that $(\eact, f_j) \in R_S$. Then $\eact$ must stem from the trace. We then apply repair (\ref{itm:repair:remove:act}) to remove an activation event.
\end{compactenum}
It can be checked that in all cases, the invariants $(i)-(iv)$ remain satisfied.
\end{compactenum}

This concludes the proof of existence of a sequence $S_0, S_1, S_2, \dots$.

\smallskip
% need lex! violations can increase
For  $S = \tup{\E,C}$ a state in this sequence, consider the measure $M(S)=(m - |R_S|, traceEvents(\E), bnd(\E), viol(S))$, where $|R_S|$ is the number of pairs in $R_S$, $traceEvents(\E)$ is the number of events in $\E$ that stem from the trace, $bnd(\E)$ the number of bound variables in events in $\E$ stemming from the trace, and $viol(S)$ the number of violations in $S$.

We observe that along the sequence $S_0, S_1, S_2, \dots$, we have $M(S_0) > M(S_1) > M(S_2) > \dots$, where we compare tuples lexicographically: Indeed, the measure decreases for repairs where an event was added because we always add an entry to $R_S$; for all other repairs, $R_S$ stays the same. The measure decreases when removing an event as the number of trace events decreases; while for all other repairs the number of trace events stays the same. When freeing an attribute, the number of bound variables decreases, and when enforcing constraints, the number of violations decreases (while the number of bound variables is unaffected).

Thus by well-foundedness, we must at some point reach a state $S_g=(\E,C)$ with $viol(S_g)=0$, i.e., a goal state.
We show that $m = |R_{S_g}|$: Suppose
to the contrary that there is an event $f_j$ in $\run$ with no $e$ such that $(e, f_j ) \in R_{S_g}$.
There must be a move with $f_j$ in $\gamma$, but it cannot be an edit or synchronous move by condition $(ii)$.
So it must be a model move. Since $\mu$ satisfies $C \setminus \{C_0\}$ restricted to $E|_{R_{S_g}}$, and $S_g$
has no violation, so with Lem.~\ref{lem:alignment:model} we conclude that also $\tup{f_1,\dots, f_{j-1}, f_{j+1}, \dots, f_m}$ must satisfy $\MM$,  which contradicts minimality of $\gamma$.
Hence $m=|R_{S_g}|$, so every event in $\run$ has a matching event in $\E$.
By assumption $(iii)$, assignment $\mu$ satisfies all constraints in $C$.
Hence $\mu$ can be used in \Cref{alg:extract-alignment} to obtain an alignment of $\trace$ and $\MM$ that is equivalent to $\gamma$.
\qed
\end{proof}

\correctness*
\begin{proof}
By Lem.~\ref{lem:soundness}, every alignment extracted from a goal state is an alignment for $\trace$ and $\MM$.
Moreover, by Lem.~\ref{lem:completeness}, for an optimal alignment $\gamma$ of $\trace$ and $\MM$ there is some goal state $S_g$ that allows to extract an alignment equivalent to $\gamma$, and by Lem.~\ref{lem:soundness} it satisfies $cost(S)=\kappa(\gamma)$. 
The claim then follows from correctness of A$^*$, i.e., the fact that a state with minimal cost is returned.
\qed
\end{proof}

}

\end{document}